\documentclass[12pt]{article}
\usepackage{graphicx}
\usepackage{amsmath}
\usepackage{amssymb}

\begin{document}

\begin{titlepage}
\begin{flushright}
OSU-HEP-00-06\\
UTEXAS-HEP-00-12\\
\end{flushright} \vskip 2cm
\begin{center}
{\Large\bf Collider Implications of Kaluza-Klein \\
Excitations of the Gluons} \vskip 1cm {\normalsize\bf
D.A.\ Dicus,$\,{}^{(a)}$\footnote{email:
phbd057@utxvms.cc.utexas.edu} C.D.\
McMullen$\,{}^{(b)}$\footnote{email:  mcmulle@okstate.edu} and S.\
Nandi$\,{}^{(b)}$\footnote{email:
shaown@osuunx.ucc.okstate.edu}\addtocounter{footnote}{-3}
\\} \vskip 0.5cm
{\it ${}^{(a)}\,$Center for Particle Physics, University of Texas, \\
Austin, Texas  78712, USA\\ [0.1truecm]
${}^{(b)}\,$Department of Physics, Oklahoma State University\\
Stillwater, OK~~74078, USA\\[0.1truecm]
}

\end{center}
\vskip 2.5cm

\begin{abstract}

We consider an asymmetric string compactification scenario in
which the SM gauge bosons can propagate into one TeV$^{-1}$-size
extra compact dimension.  These gauge bosons have associated KK
excitations that present additional contributions to the SM
processes.  We calculate the effects that the KK excitations of
the gluons, $g^{\star}$'s, have on multijet final state production
in proton-proton collisions at the Large Hadron Collider energy.
In the case of dijet final states with very high $p_{{}_T}$, the
KK signal due to the exchanges of the $g^{\star}$'s is several
factors greater than the SM background for compactification scales
as high as about 7 TeV.  The high-$p_{{}_T}$ effect is not as
dramatic for the direct production of a single on-shell
$g^{\star}$, which subsequently decays into $q$-$\bar{q}$ pairs,
where the KK signal significantly exceeds the SM three-jet
background for compactification scales up to about 3 TeV.  We also
present our results for the four-jet final state signal from the
direct production of two on-shell $g^{\star}$'s.

\end{abstract}

\end{titlepage}

\newpage

\noindent
{\bf 1.  Introduction}

\vspace{0.2cm}

\noindent Recent developments in superstring theory have sparked
much interest in scenarios where the string scale is much smaller
than the four-dimensional Planck scale~\cite{superstring}.  The
size of the six extra compact dimensions may be much larger than
the inverse Planck scale, giving rise to many new phenomenological
possibilities. For $n$ large extra dimensions compactified at the
same scale $R^{-1}$, the size $R$ is related to the
four-dimensional Planck scale $M_P$ via the relation

\vspace{-8pt} \begin{equation} \label{eq:Planck} M_{P}^2 =
M_{\star}^{n+2} \, R^n \, ,
\end{equation}

\noindent where $M_{\star}$ is the ($4 \! + \! n$)-dimensional
Planck scale, which is of the order of the string scale. Recently,
it was shown that this relation (\ref{eq:Planck}) is
phenomenologically viable~\cite{Planck} for $n \geq 2$, $R$ can be
in the sub-millimeter regime, and the string scale could be fairly
close to the electroweak scale, namely, a few tens of a TeV.  The
gauge hierarchy problem is eliminated since the four-dimensional
Planck scale $M_P$ is not a fundamental quantity in this scheme.
If all six extra dimensions from the superstring theory are
compactified at the same scale, then $1/R$ is about 10 MeV.  Thus,
if the Standard Model (SM) particles are allowed to propagate into
these extra dimensions (the bulk), they will have Kaluza-Klein
(KK) excitations with masses at the 10 MeV scale.  The
non-observation of such KK states up to about a TeV at present
high-energy colliders therefore implies, in such scenarios, that
all SM particles are confined to a three-dimensional brane (D$_3$
brane) of the usual three spatial dimensions.  These are the key
features of the class of models based on the symmetrical
compactification proposal of Arkani-Hamed, Dimopoulos, and Dvali
(ADD)~\cite{Planck} for solving the hierarchy problem.

It is also possible, however, to devise a model with asymmetrical
compactification where SM particles live in a brane which extends
into one or more TeV$^{-1}$-size extra dimensions.  The lowest
lying KK excitations then have masses at the TeV scale, at the
edge of the grasp of present high-energy colliders.  Such a scheme
has many interesting consequences.  For example, it alters the
evolution of the gauge couplings from the usual logarithmic to
power law behavior~\cite{unify}.  The unification scale can be
several orders of magnitude smaller~\cite{unify}, even as low as a
few TeV. Recently, an asymmetrical compactification scenario was
proposed with two distinct compactification scales~\cite{asym}:
$n$ dimensions of size $R \sim$ mm and $m$ of size $r \sim$
TeV$^{-1}$.  In particular, we consider the $n=1, \, m=5$ case.
The scaling relation for this model is~\cite{asym}

\vspace{-8pt} \begin{equation} M_{P}^2   =   M_{\star}^3  \, R \,
=  \, M^8  \,  R  \,  r^5 \, .
\end{equation}

\noindent It was shown in Ref.~\cite{asym} that this model
satisfies all of the current astrophysical and cosmological
constraints~\cite{astro}. With $1/R \sim 10^{-3}$ eV and $1/r \sim
1$ TeV, we get $M \sim 100$ TeV and $M_{\star} \sim 10^5$ TeV. In
this scenario, the SM gauge bosons (and perhaps the Higgs boson)
can propagate into one of the TeV$^{-1}$-size extra dimensions,
while the SM fermions are confined to the usual D$_3$ brane.  $M
\sim 100$ TeV is then consistent with the unification scale
(assuming about a factor of ten uncertainty due to threshold and
other effects). The smoking gun signatures of this scenario are
deviations from Newton's law of gravity in the sub-millimeter
regime as well as new high-$p_{{}_T}$ jet physics in high-energy
hadron colliders.

Most of the work on the collider phenomenology of extra
dimensions~\cite{collider} has been on the ADD scenario in which
only the graviton propagates in the bulk. Hence, the only
additional contribution to collider processes stems from the KK
excitations of the graviton.  The contributions of individual KK
modes, with $4$D gravitational strength, to collider processes is
extremely small. However, the compactification scale {$\mu$} is so
small ($\mu \sim $ mm$^{-1} \sim 10^{-3}$ eV) that a very large
number of such modes contribute in a TeV-scale collider process,
yielding a significant total deviation from the SM results.
Studies of various collider processes typically give a bound on
the string scale (taken approximately to be the cut-off scale) of
about a TeV~\cite{collider}.

The asymmetric scenario, in which SM fields, in addition to
gravity, may propagate in one or more extra dimensions of
TeV$^{-1}$-size, will have a more direct effect in high-energy
collider processes.  Beginning with the original suggestion by
Antoniadis~\cite{Antoniadis}, some work has also been done for the
collider phenomenology of this scenario~\cite{asymcoll}, including
the effects on EW precision measurements~\cite{ew}, Drell-Yan
processes in hadronic colliders~\cite{muon}, and $\mu^{+} \mu^{-}$
pair production in electron-positron colliders~\cite{muon}.  The
typical bound is 1--2 TeV for the compactification scale.

In this work, we study the scenario proposed in Ref.~\cite{asym},
in which only the SM gauge bosons (and perhaps the Higgs boson)
propagate into one of the TeV$^{-1}$-size extra
dimensions.\footnote{However, our results apply to any
compactified string model in which the gluons propagate into one
such extra dimension.}\addtocounter{footnote}{-1} More
specifically, we study the effects that the KK excitations of the
gluons have on multijet production in high-energy hadronic
colliders such as the Large Hadron Collider (LHC).  We calculate
the modifications to the SM cross sections for multijet final
states which arise from the direct production and exchanges of KK
excitations of the gluons. At the LHC energy, we find substantial
deviations from the SM predictions for dijet final states up to a
compactification scale of about 7 TeV; whereas for the Tevatron,
the KK contribution only exceeds the SM background for small
compactification scales ($\lesssim 2.0$ TeV). For the direct
production of a $g^{\star}$ on-shell at the LHC, which
subsequently decays into $q$-$\bar{q}$ pairs, the effect is not as
pronounced as the dijet case, but is still significant.  We also
present the contribution of the production of two on-shell
$g^{\star}$'s.  Our paper is organized as follows. We briefly
discuss our formalism in Section 2, and supplement this with
additional details in the Appendix.  In Section 3, we calculate
the effects that the exchanges of $g^{\star}$'s have on dijet
production and discuss our results and the significance of the SM
background. Our analytic expressions for the cross sections for
the processes leading to the direct production of one or two
on-shell $g^{\star}$'s are presented in Sections 4 and 5,
respectively; also included are a discussion of our numerical
results and, for the single $g^{\star}$ case, comparison to the SM
three-jet background.  Section 6 contains our conclusions.

\vspace{0.5cm}

\pagebreak[4]
\noindent
{\bf 2.  Formalism}

\vspace{0.2cm}

\noindent We are interested in tree-level parton subprocesses
involving the exchanges or direct production (or both) of KK
excitations of gluons. The starting point is the generalization of
the $4$D SM Lagrangian density to the $5$D Lagrangian density.
Integration over the fifth dimension then yields the effective
$4$D Lagrangian density, which includes the usual $4$D SM
Lagrangian density plus terms involving the KK excitations of the
SM gauge fields.  These KK terms dictate the possible couplings
that the KK excitations can have both with each other and with the
SM fields, and provide the Feynman rules for these vertices as
well as the KK propagators.

In the model under consideration, the SM gauge bosons can
propagate into one large extra compact dimension.  The terms in
the $5$D Lagrangian density relevant to us are (1) the terms
involving the contraction of the $5$D gluon field strength tensors
$F_{\mathit{MN}}^a  =  \partial_M A_N^a - \partial_N A_M^a -
g_{{}_5} f^{\mathit{abc}} A_{\mathit{M}}^b A_{\mathit{N}}^c$ with
$5$D indices $M,N \in \{0,1,...,4\}$, where $g_{{}_5}$ is the $5$D
strong coupling and $a$,$b$,$c$ are the usual gluon color indices;
and (2) the terms involving the quark fields, which contain a
delta function to constrain the SM fermions to the D$_3$ brane:

\vspace{-3pt} \begin{equation} \mathcal{L}_5  =
-\frac{1}{4}F_{\mathit{MN}}^a F^{\mathit{MNa}} + i \bar{q}
\gamma^{\mu} D_{\mu} q \delta (y) \, .
\end{equation}

\noindent Here, $D_{\mu}$ is the usual $4$D covariant derivative,
$\mu$,$\nu$ are the usual $4$D spacetime indices, and the
compactified extra dimension coordinate $y$ is related to the
radius of the extra dimension $r$ by $y  =  r \phi$. We consider
compactification on a $S^1 / Z_2$ orbifold with the orbifold
symmetry, $y \rightarrow -y$ such that $A_{\mu}^a (x,-y) =
A_{\mu}^a (x,y)$, and impose the gauge choice $A_{4}^a (x,y) = 0$.
This is the unitary gauge.  The $5$D gluon field $A_{\mu}^a (x,y)$
can then be Fourier expanded in terms of the compactified
dimension $y$ as

\vspace{-3pt} \begin{equation} A_{\mu}^a (x,y) =
\frac{1}{\sqrt{\pi r}}\mbox{\raisebox{-.6ex}{\huge $[$}}A_{\mu
0}^a (x) + \sum_{n=1}^{\infty}A_{\mu n}^a (x)
\cos(n\phi)\mbox{\raisebox{-.6ex}{\huge $]$}} \, ,
\end{equation}

\noindent where the normalization of $A_0^a (x)$ is one-half that
of $A_n^a (x)$. When the $5$D Lagrangian density is integrated
over the extra dimension $y$, this sum represents a tower of KK
excitations $A_{\mu n}^a (x)$ of the gluon field.  The $n  = 0$
mode gluon is identified with the observed massless gluon of the
SM, denoted by $g$, while the $n  >  0$ KK modes, denoted by
$g_n^{\star}$, have masses $m_n  =  n \, \mu$ where $\mu$ is the
compactification scale ($1/r$). It will prove convenient to refer
to the $n  =  0$ and $n  >  0$ modes separately by letting
``gluon'' or $g$ represent just the $n  =  0$ mode, and letting
``KK excitation of the gluon'' or $g^{\star}$ or $g_n^{\star}$
strictly imply $n  >  0$.

\begin{figure}
\setlength{\abovecaptionskip}{10pt}
\centering{\includegraphics[bb=85.5 171 517.5 675]{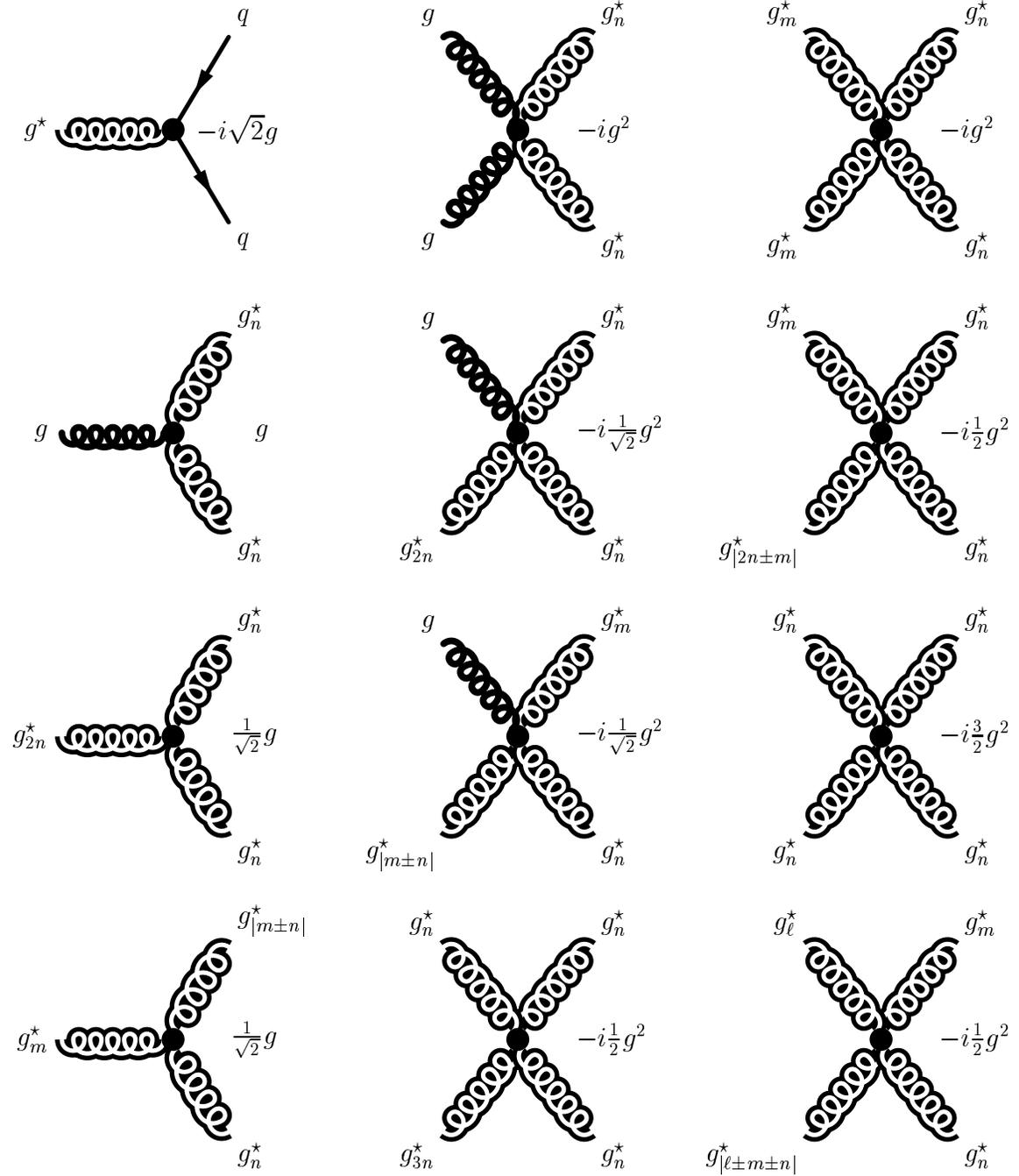}}
\caption{Relative coupling strengths of vertices involving
$g^{\star}$'s. Only the overall factors are shown:  The
$q$-$\bar{q}$-$g^{\star}$ vertex also involves the SU($3$) matrix
element and the Dirac $\gamma_{\mu}$ matrix; triple vertices of
$g$'s and $g^{\star}$'s also include the usual SU($3$) structure
functions and the momenta factors; and quadruple vertices of $g$'s
and $g^{\star}$'s also contain the usual structure function
factors as well as the metric tensors $g_{\mu \nu}$. Here, $n$,
$m$, and $\ell$ are distinct positive integers ($n  \neq  m  \neq
\ell$).} \label{fig:FDgstar} \setlength{\abovecaptionskip}{10pt}
\end{figure}

The detailed procedure for integrating over the fifth dimension
$y$ to obtain, in the effective $4$D theory, the factors for the
allowed vertices involving KK excitations of the gluons may be
found in the Appendix, and lead to the coupling strengths
displayed in Fig.~\ref{fig:FDgstar}. Notice that a single
$g^{\star}$ can couple to quarks, but not to gluons. Furthermore,
quark-less vertices with $N$ $g^{\star}$'s only have non-vanishing
coupling strengths if the modes
$n_{{}_{1}}$,$n_{{}_{2}}$,$\ldots$,$n_{{}_N}$ of the $g^{\star}$'s
satisfy the relation
\vspace{-8pt} \begin{equation} \label{eq:modes} \mid \! n_{{}_{1}}
\, \pm \, n_{{}_{2}} \, \pm \, \cdots \, \pm \, n_{{}_{N-1}} \!
\mid \,= n_{{}_N} \, .
\end{equation}

\noindent Although this relation, Eq.~(\ref{eq:modes}), governs
the possible vertices, it is not a law expressing $5$D momentum
conservation for $N \rightarrow M$ processes:  For example, a
$g^{\star}$ can not decay into gluons at the tree level, although
this process is permitted when a quark loop is introduced.  Also
worth noting are the factors of $\sqrt{2}$, which originate from
the different rescaling of the $n =  0$ and $n  >  0$ modes,
necessary to obtain canonically normalized kinetic energy terms in
the effective $4$D Lagrangian density~\cite{rescale}.

Another difference between the Feynman rules for the $g$ and the
$g^{\star}$ lies in the propagator. The $g^{\star}$ propagator is
that of a usual massive gauge boson, shown here in the unitary
gauge:

\vspace{-3pt} \begin{equation} \label{eq:prop} - i \Delta_{\mu\nu
n}^{ab}(p^2) = -i \delta^{ab} \frac{g_{\mu \nu}  - \frac{p_{\mu}
p_{\nu}}{m_n^2}}{p^2  -  m_n^2  +  i m_n \Gamma_n} \, .
\end{equation}

\noindent At tree-level, the $g_n^{\star}$ decays into $q  \bar{q}
$ pairs with (total) width $\Gamma_n  =  2 \, \alpha_{{}_S} (Q)
m_n$.\footnote{We neglect the top quark mass relative to the very
heavy $g^{\star}$.}\addtocounter{footnote}{-1} The decay width can
not be neglected because the subprocess energy $\sqrt{ \hat{s} }$
runs up to $14$ TeV at the LHC, while we are interested in
TeV-scale compactification. For diagrams where a virtual $g$ or
$g^{\star}$ exchanges between two quark pairs (\textit{e.g.}, in
$q \bar{q} \rightarrow  q \bar{q} $), there is the usual diagram
with the $g$ propagator in addition to a tower of diagrams with
$g_n^{\star}$ propagators, or, equivalently, an effective
propagator given by the sum

\vspace{-3pt} \begin{equation} \Delta_{\mathit{eff}}(p^2)  =
c_0\Delta_0 (p^2) \, +  \, \sum_{n=1}^{\infty} c_n\Delta_n(p^2) \,
.
\end{equation}

\noindent Notice that $c_n$ incorporates the different
$q$-$\bar{q}$-$g$ and $q$-$\bar{q}$-$g_n^{\star}$ vertex factors
(\textit{i.e.}, $c_0  =  1$, $c_{n>0}  =  2$).  This effective
propagator can be generalized to the case of arbitrary vertices
with appropriate choices of the $c_n$ factors (including setting
$c_n$ equal to zero when either vertex is forbidden).

The mass of the $g^{\star}$ also enters into the expression for
the cross section via summations over polarization states when
external $g^{\star}$'s are present. For the direct production of
$g^{\star}$'s, the summation of polarization states is given by

\vspace{-3pt} \begin{equation} \sum_{\sigma} \epsilon_{\mu
n}^{a\ast}(k,\sigma)\epsilon_{\nu n}^b(k,\sigma)  =
\mbox{\raisebox{-.6ex}{\huge $($}} \!\! -g_{\mu \nu}
 +  \frac{k_{\mu}k_{\nu}}{m_n^{2}}\mbox{\raisebox{-.6ex}{\huge
$)$}}\delta^{ab} \, .
\end{equation}

\noindent
Compare this to the case of external $g$'s,
in which case a projection such as

\vspace{-3pt} \begin{equation} \sum_{\sigma}
\epsilon_{\mu}^{a\ast}(k,\sigma)\epsilon_{\nu}^b(k,\sigma)  =
\mbox{\raisebox{-.6ex}{\huge $[$}} \! -g_{\mu \nu} +
\frac{(\eta_{\mu}k_{\nu}  +  \eta_{\nu}k_{\mu})}{(\eta\cdot k)} -
\frac{\eta^2 k_\mu k_\nu}{(\eta\cdot
k)^2}\mbox{\raisebox{-.6ex}{\huge $]$}}\delta^{ab}
\end{equation}

\noindent can be made to eliminate unphysical longitudinal
polarization states (and thereby satisfy gauge invariance), for
arbitrary four-vector $\eta_{\mu}$.

\vspace{0.5cm}

\noindent {\bf 3.  Dijet Production}

\vspace{0.2cm}

\noindent For dijet production, all tree-level diagrams are
included which do not contain any $g^{\star}$'s in the final
state, since the $g^{\star}$'s would quickly decay into $q \bar{q}
$ pairs, thereby producing additional jets.\footnote{We neglect
the contributions from cases where multiple jets are produced, but
only two of them pass the various cuts.} Thus, the KK excitations
only appear in two-jet diagrams via virtual $g^{\star}$
propagators. The net tree-level effect of the $g^{\star}$'s on
dijet production is the replacement of the SM gluon propagator by
an effective KK propagator, wherever five-momentum is conserved.
Employing gauge invariance, we drop the second term in
Eq.~(\ref{eq:prop}) in our analysis of dijet production.
It is then convenient to define $D_n(p^2)$ and
$D_{\mathit{eff}}(p^2)$ as

\vspace{-8pt} \begin{eqnarray}
D_n(p^2) &\!\! = &\!\! \frac{c_n}{p^2  -  m_n^2  +  i m_n \Gamma_n} \nonumber \\
D_{\mathit{eff}}(p^2) &\!\! = &\!\! \frac{c_0}{p^2} \, +  \,
\sum_{n=1}^{\infty} c_n D_n(p^2) \, .
\end{eqnarray}

\noindent Here, $c_n$ represents the fact that the
$q$-$\bar{q}$-$g$ and the $q$-$\bar{q}$-$g_n^{\star}$ vertex
factors differ by a $\sqrt{2}$ (\textit{i.e.}, $c_0  =  1$,
$c_{n>0} \, = \, 2$). In the amplitude-squared, it is therefore
necessary to evaluate terms of the form

\begin{figure}
\centering{\includegraphics[bb=108 171 513 648]{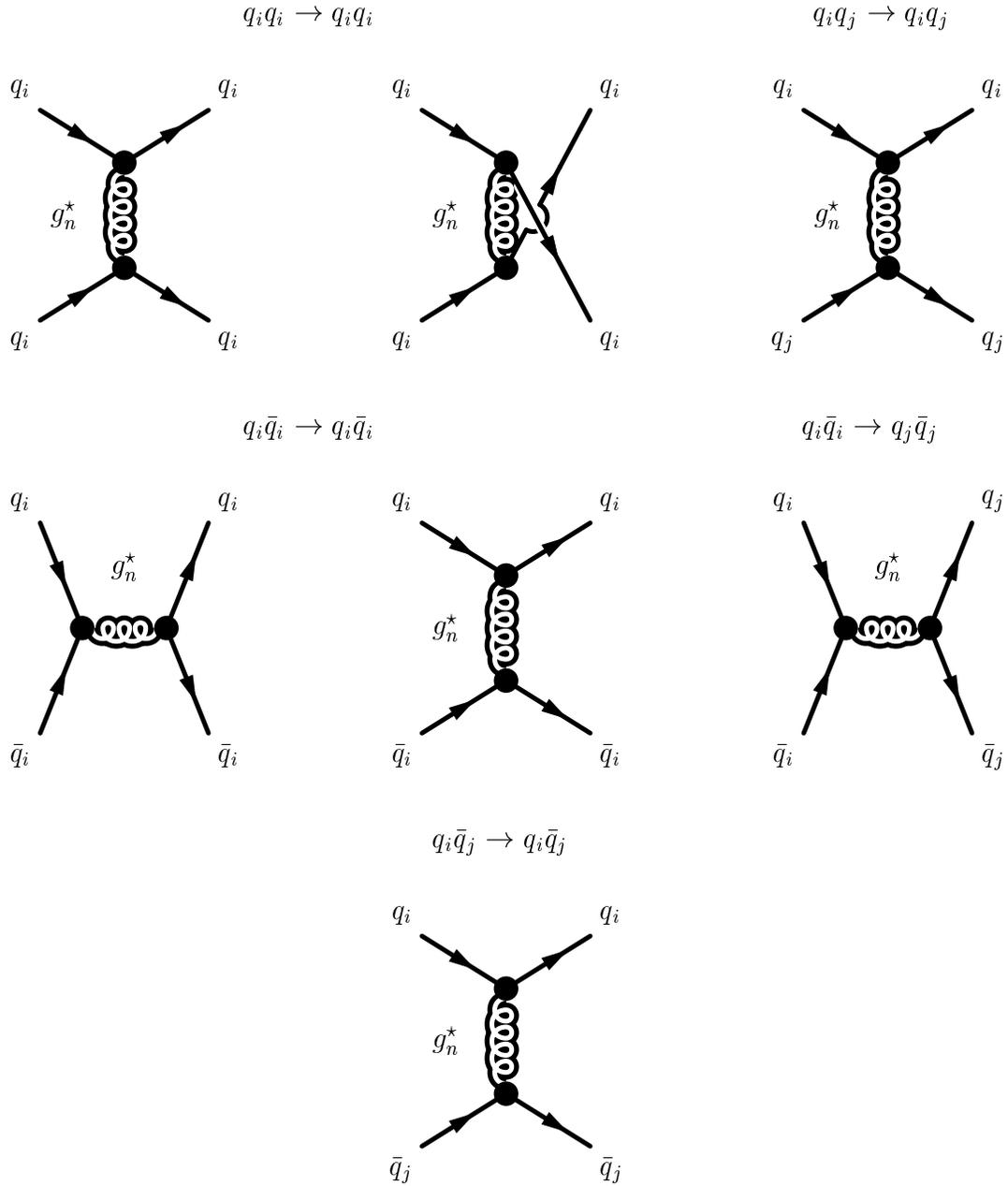}}
\caption{Dijet diagrams involving KK excitations of the gluons.
The indices $i$ and $j$ represent distinct ($i \neq j$) quark
flavors.} \label{fig:2jets}
\end{figure}

\vspace{-3pt} \begin{equation} \label{eq:Deff}
\frac{1}{2}\mbox{\raisebox{-.6ex}{\huge
$[$}}D_{\mathit{eff}}^{\star}( \hat{v} )D_{\mathit{eff}}( \hat{w}
)
 +  D_{\mathit{eff}}( \hat{v} )D_{\mathit{eff}}^{\star}(
\hat{w} )\mbox{\raisebox{-.6ex}{\huge $]$}}
  =  \sum_{m,n=0}^{\infty} c_m c_n \frac{ \hat{v} '_m  \hat{w} '_n  +  m_m \Gamma_m m_n \Gamma_n}{(v_m^{2}  +
m_m^{2}\Gamma_m^{2}) (w_n^2  +  m_n^2\Gamma_n^2)} \, ,
\end{equation}

\noindent where $ \hat{v} $ and $ \hat{w} $ are any of the three
usual (subprocess) Mandelstam variables (\textit{i.e.}, $ \hat{v}
, \hat{w} \in \{  \hat{s} , \hat{t} , \hat{u}  \}$), and $ \hat{v}
'_n$ represents the subtraction of $m_n^2$ from $\hat{v}$
(\textit{i.e.}, $ \hat{v} '_n  \equiv   \hat{v} _n-m_n^2$). (In
Eq.~(\ref{eq:Deff}) we make an exception and include the $n = 0$
and $n > 0$ modes together for conciseness.)  This sum converges
somewhat rapidly:\footnote{When generalizing to the case where the
gluons may propagate into more than one large extra dimension, the
sum in the effective propagator is formally divergent.  However,
this problem has been widely addressed in the
literature~\cite{convergence}, where various solutions have been
proposed.} Since $ \sqrt{\hat{s}} $ runs up to $14$ TeV for the
LHC, the sum can be truncated after a couple dozen terms
(\textit{i.e.}, when $n$ becomes at least a couple of times
greater than $14$ TeV$\, / \mu$, where $\mu$ is the
compactification scale). We choose $n_{{}_{\mathit{max}}} = 50$.
From five-momentum conservation, there are no internal
$g^{\star}$'s for any tree-level dijet diagrams involving external
gluons (\textit{e.g.}, the KK excitations do not affect the
process $q \bar{q}   \rightarrow gg$).  The diagrams to which the
KK excitations do contribute are illustrated in
Fig.~\ref{fig:2jets}.

The total dijet cross section $\sigma(p p  \rightarrow  2 \,
\mathrm{jets})$ is obtained from the individual subprocess cross
sections $\hat{\sigma}(ab  \rightarrow  cd)$ and the parton
distributions $f_{a/{}_A}(x_{{}_A},Q)$ and
$f_{b/{}_B}(x_{{}_B},Q)$ by integrating over the momentum
fractions $x_{{}_A}$ and $x_{{}_B}$ and summing over all possible
subprocesses $ab  \rightarrow  cd$:

\vspace{-3pt} \begin{equation} \sigma(pp  \rightarrow  2 \,
\mathrm{jets})  = \sum_{ab\rightarrow
cd}\,\,\,\int_{\mbox{\raisebox{-1.5ex}{\scriptsize{$\!\!\!\!\!\!\!\!
4 p_{{}_T}^2/ s $}}}}^{\mbox{\raisebox{.9ex}{\scriptsize{$\!\!\!
1$}}}} d\tau\frac{d{\mathcal{L}}}{d\tau}\hat{\sigma}(ab
\rightarrow cd) \, .
\end{equation}

\noindent Here, $p_{{}_T}$ is the transverse momentum and
$d\mathcal{L}/d\tau$ is the parton luminosity:

\vspace{-3pt} \begin{equation} \frac{d{\mathcal{L}}}{d\tau}  =
\int_{\mbox{\raisebox{-1ex}{\scriptsize{$\!\!\!\!
\tau$}}}}^{\mbox{\raisebox{1ex}{\scriptsize{$\!\!\!
1$}}}}\frac{dx_{{}_A}}{x_{{}_A}} f_{a/{}_A}(x_{{}_A},Q)
f_{b/{}_B}(x_{{}_B},Q) \, .
\end{equation}

\begin{figure}
\setlength{\abovecaptionskip}{0pt}
\centering{\includegraphics[bb=148.5 153 463.5 630]{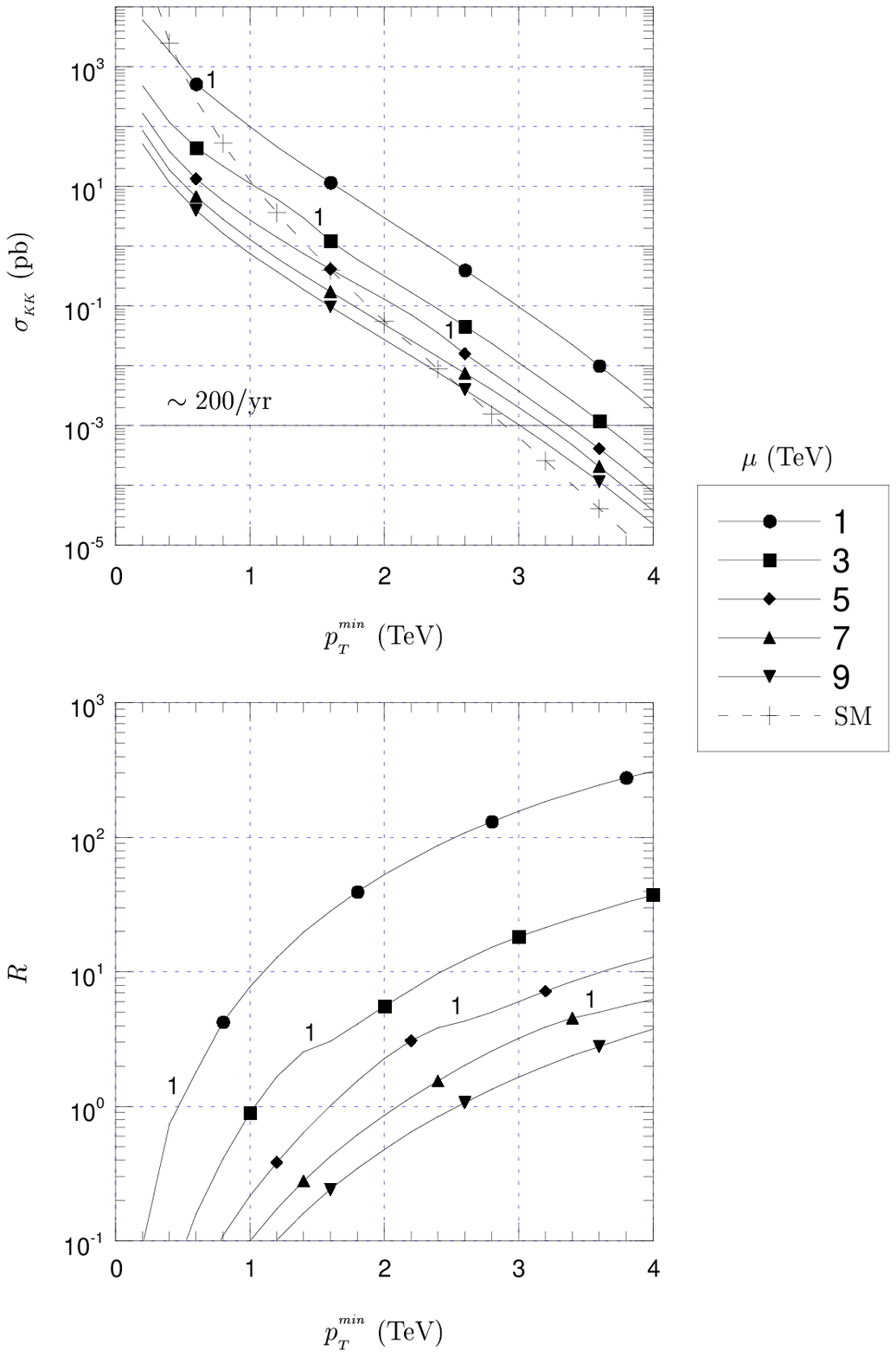}}
\vspace{10pt} \caption{The contributions of the virtual exchanges
of $g^{\star}$'s to the LHC dijet production cross section,
$\sigma_{{}_{\mathit{KK}}}  =  \sigma -
\sigma_{{}_{\mathit{SM}}}$, (top) and the ratio of the KK
contribution to the SM background, $R  = \sigma_{{}_{\mathit{KK}}}
/ \sigma_{{}_{\mathit{SM}}}$, (bottom) are illustrated as a
function of the minimum transverse momentum
$p_{{}_T}^{{}^{\mathit{min}}}$ for fixed values of the
compactification scale $\mu$. The solid horizontal line represents
$\sim 200$ events/yr at the projected integrated luminosity.
Discernible bumps in regions for which
$p_{{}_T}^{{}^{\mathit{min}}} = k \mu / 2$ are indicated by the
corresponding value of $k \in \{1,2,\ldots\}$.}\label{fig:gs2a}
\setlength{\abovecaptionskip}{0pt}
\end{figure}

\begin{figure}
\setlength{\abovecaptionskip}{0pt}
\centering{\includegraphics[bb=141 222 456 699]{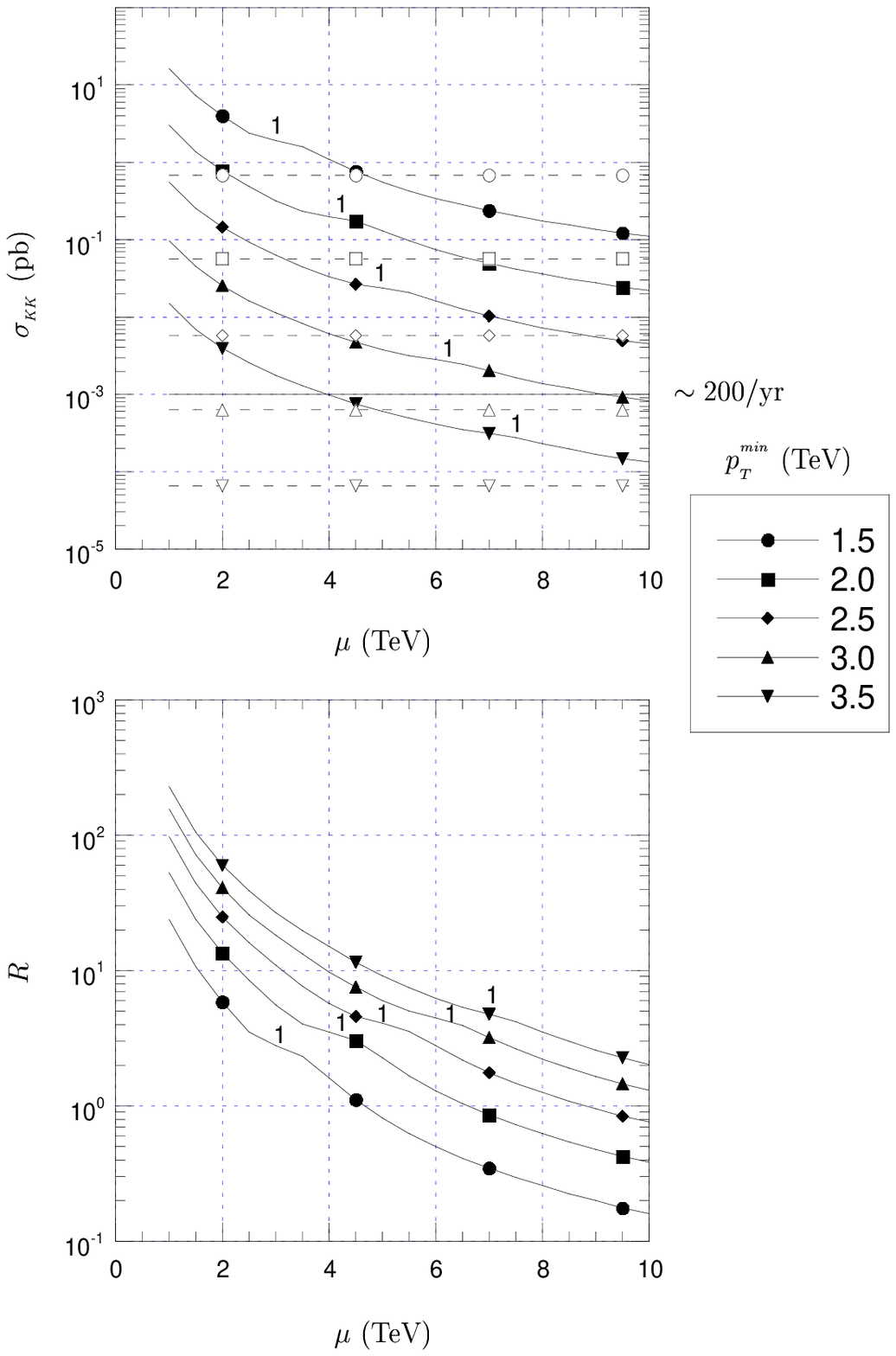}}
\vspace{10pt} \caption{Same as Fig.~\ref{fig:gs2a}, but as a
function of the compactification scale $\mu$ for fixed values of
the minimum transverse momentum $p_{{}_T}^{{}^{\mathit{min}}}$.
The horizontal dashed lines represent the SM
background.}\label{fig:gs2b}
\end{figure}

\noindent We evaluate the CTEQ distribution functions~\cite{CTEQ}
for the parton luminosity at $Q = p_{{}_T}$, and impose the
following cuts: The transverse momentum $p_{{}_T}$ is constrained
to lie above some minimum $p_{{}_T}^{{}^{\mathit{min}}}$, while
the rapidity is restricted to satisfy $\mid\! y \!\mid \, \leq
2.5$. The total cross section can also be separated into the SM
cross section and the $g^{\star}$ cross section, which is due to
the contributions of Fig.~\ref{fig:2jets}: $\sigma  =
\sigma_{{}_{\mathit{SM}}} + \sigma_{{}_{\mathit{KK}}}$.  Although
$\sigma_{{}_{\mathit{KK}}}$ includes the interference terms
between $g$'s and $g^{\star}$'s, it usefully represents the amount
by which the total cross section exceeds the SM background. The KK
contributions, along with the SM background, are shown in
Fig.'s~\ref{fig:gs2a}--\ref{fig:gs2b}
for compactification scales in the range $1$ TeV $\leq \mu \leq
10$ TeV and for transverse momentum as high as
$p_{{}_T}^{{}^{\mathit{min}}} \leq 4$ TeV.

The KK effect is actually quite large:  For sufficiently high
$p_{{}_T}^{{}^{\mathit{min}}}$ ($\sim 2$ TeV), the effect of the
virtual exchanges of the $g^{\star}$'s actually exceeds the SM
background for compactification scales below $7$ TeV.  The effect
becomes even more pronounced for yet higher
$p_{{}_T}^{{}^{\mathit{min}}}$, where the KK contribution becomes
several factors larger than the SM cross section.  The trend
continues beyond the $4$ TeV shown, but the cross section is too
small beyond this point to observe more than a couple of events
per year at the anticipated integrated luminosity of the LHC ($2
\times 10^5$ pb$^{-1}$).  Final quark states due to the decay of a
very massive $g^{\star}$ have very high $p_{{}_T}$, thereby
enhancing the ratio $R  \equiv \sigma_{{}_{\mathit{KK}}} /
\sigma_{{}_{\mathit{SM}}}$ for high
$p_{{}_T}^{{}^{\mathit{min}}}$, which is where the $g^{\star}$
contribution actually exceeds the SM contribution.  When
$p_{{}_T}^{{}^{\mathit{min}}} = k \mu / 2$ for $k \in
\{1,2,\ldots\}$, there is a slight disturbance in the cross
section plots, which is expected since this corresponds to an
on-shell $g^{\star}$ contribution.  Naturally, the disturbance is
only discernible for small values of $k$.  These discernible
regions are indicated on the plots by the corresponding values of
$k$.

\begin{figure}
\setlength{\abovecaptionskip}{0pt}
\centering{\includegraphics[bb=143 463 544 698]{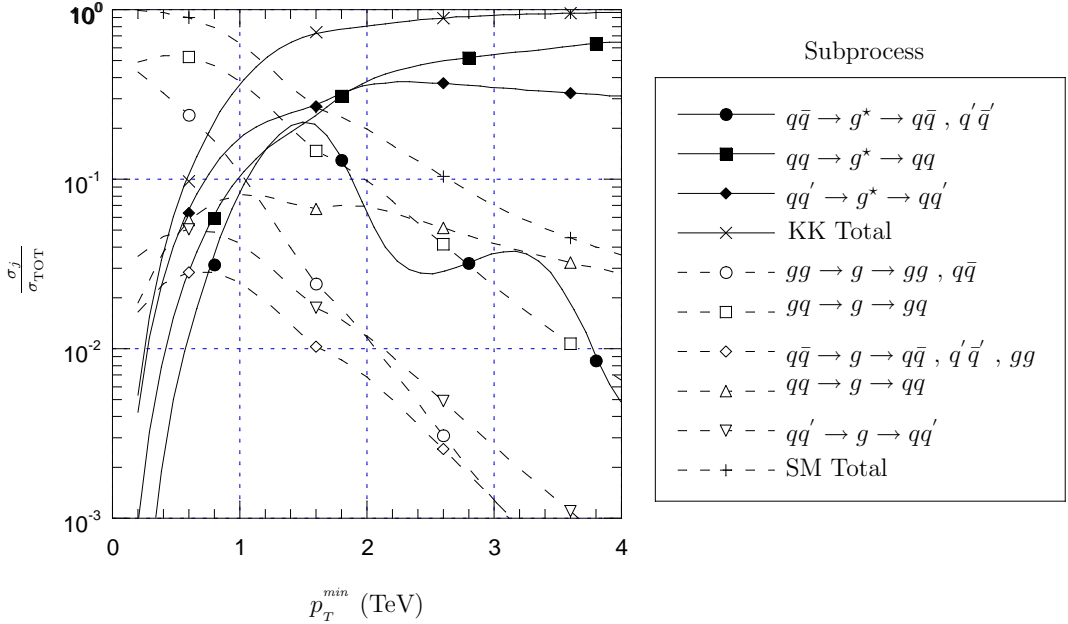}}
\vspace{0pt} \caption{The partial contributions to the total dijet
cross section are shown as a function of
$p_{{}_T}^{{}^{\mathit{min}}}$, for $\mu = 3.5$
TeV.}\label{fig:sm2p} \setlength{\abovecaptionskip}{0pt}
\end{figure}

The partial contributions of the various subprocesses to the full
dijet KK (for a representative value of $\mu = 3.5$ TeV) and SM
cross sections are illustrated in Fig.~\ref{fig:sm2p}.
At low $p_{{}_T}$, the virtual $g^{\star}$ effect is greatest for
subprocesses with two different initial quarks, while, at high
$p_{{}_T}$, it is largest for subprocesses with identical initial
quarks.

\begin{figure}
\setlength{\abovecaptionskip}{0pt}
\centering{\includegraphics[bb=171 306 490.5 531]{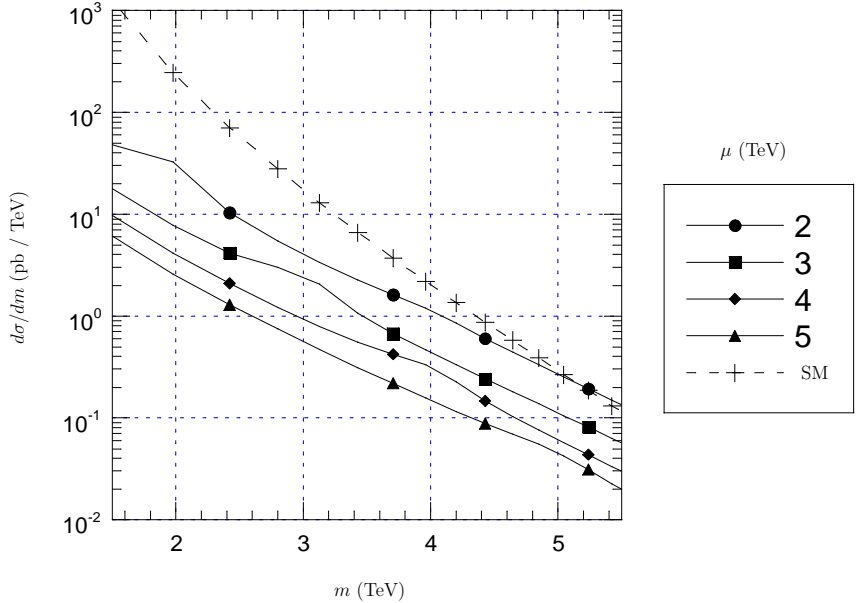}}
\vspace{10pt} \caption{The differential cross section $d\sigma / d
m$ is shown as a function of the invariant mass $m$ of the
$q$-$\bar{q}$ pair. The peaks that are predicted to occur when the
invariant mass $m$ matches the compactification scale $\mu$ are
subtle and located well below the SM signal. } \label{fig:fdsdo}
\setlength{\abovecaptionskip}{0pt}
\end{figure}

Fig.~\ref{fig:fdsdo} shows the dijet differential cross section
$d\sigma / d m$ as a function of the invariant mass $m$ of the
final state $q$-$\bar{q}$ pair:  The peaks are subtle, and
positioned well below the SM background.  The signal in the
two-jet invariant mass distribution is well below the SM
background unless the invariant mass is very large ($m > 5$ TeV).
However, at the LHC, the cross sections are not large enough for
the signal to be observable in this range of $m$.  There are two
reasons why the dijet invariant mass distribution does not give a
good signal.  First, the widths of the $g^{\star}$'s are large
such that the peaks corresponding to $m = \mu$ are not sharp nor
tall enough.  Secondly, most of the cross section for a given
invariant mas comes from pairs which have relatively low $p_T$ for
which the SM background is very large.  The decay of the resonant
KK gluon, $g^{\star}$, gives rise to high $p_T$ for each of the
jet pairs.  It is only when we consider the final states where
each of the jets have high $p_T$ that the KK contributions exceed
the SM background.  In the invariant mass distribution, such high
$p_T$ contributions constitute only a very small part of the cross
sections observable at the LHC energy.

\begin{figure}
\setlength{\abovecaptionskip}{0pt}
\centering{\includegraphics[bb=138 460 537 698]{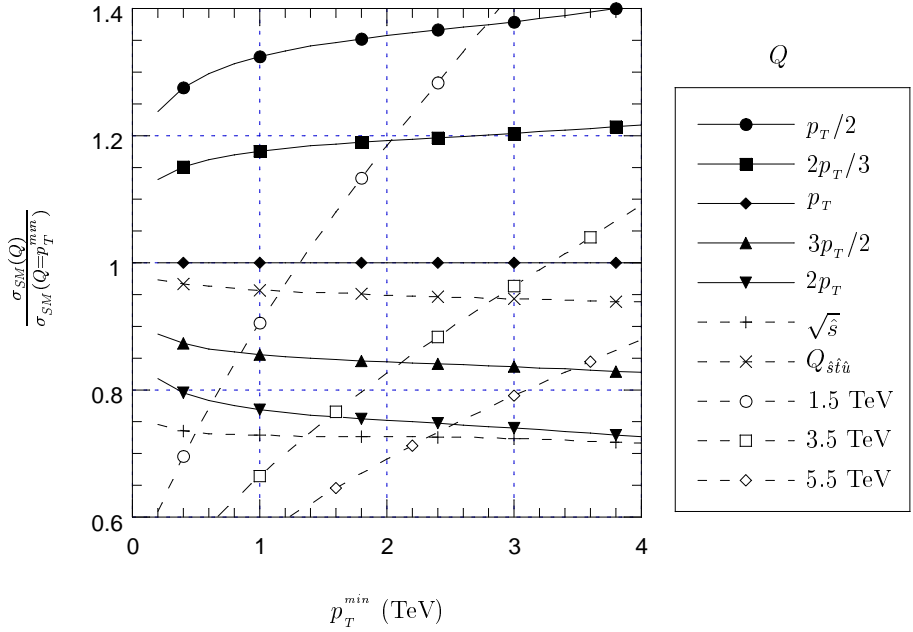}}
\vspace{10pt} \caption{The effect that variation of the choice of
$Q$ has on the SM dijet background is shown as a function of the
minimum transverse momentum, $p_{{}_T}^{{}^{\mathit{min}}}$. Here
$Q_{\hat{s}\hat{t}\hat{u}} =
\sqrt{\frac{\hat{s}\hat{t}\hat{u}}{\hat{s}^2  +  \hat{t}^2  +
\hat{u}^2}}$, and values in TeV (\textit{e.g.}, $3.5$ TeV)
correspond to the choice of (constant) $Q$ equal to a
compactification scale at that particular scale.} \label{fig:sm2q}
\setlength{\abovecaptionskip}{0pt}
\end{figure}

Depicted in Fig.~\ref{fig:sm2q} are the effects produced by
variation of the somewhat arbitrary choice of $Q  =
p_{{}_T}^{{}^{\mathit{min}}}$ for the SM background. The relative
uncertainty in the SM background can be quite high, say $40$ \%,
due to the ambiguity in the choice of $Q$, and other factors such
as the choice of parton distributions. However, since the signal
and the background are each calculated at tree-level, the
uncertainties should somewhat cancel in the ratio, $R$.  Thus, $R$
provides a good measure of the relative KK effect. We point out
that due to these uncertainties and the fact that one can not
directly measure $R$, when working at tree-level it is necessary
to look for signals that disagree with the SM by much more than
$50\%$, probably as much as $100\%$, to be sure that we are indeed
observing a signal for new physics. Therefore, the detection of KK
excitations of the gluons is most favorable for regions of
($p_{{}_T}^{{}^{\mathit{min}}},\mu$)-space where the KK
contribution is at least comparable to the SM background, and
above the horizontal line (in
Fig.'s~\ref{fig:gs2a}--\ref{fig:gs2b}) that marks an anticipated
couple of hundred events per year.

For comparison, in Fig.'s~\ref{fig:tgs2a}--\ref{fig:tgs2b} we also
give the $g^{\star}$ cross section and its relation to the SM
background for the Fermilab Tevatron $p\bar{p}$ collider running
at $\sqrt{s} =  2$ TeV. The KK effect is much smaller than for the
LHC because of the considerably more restrictive constraints on
the transverse momentum.  The $g^{\star}$ cross section is only
comparable to the SM for compactification scales $\mu$ as high as
about $2$ TeV, and the relative uncertainty in the total dijet
cross section must be quite precise in order to see a sizeable
discrepancy for $\mu \sim 3$ TeV.

\vspace{0.5cm}

\pagebreak[4] \noindent {\bf 4.  Single On-Shell $g^{\star}$
Production}

\vspace{0.2cm}

\noindent Three-jet KK final states
predominantly\addtocounter{footnote}{-2}\footnote{The
contributions of virtual $g^{\star}$ exchanges for which no
external on-shell $g^{\star}$'s are produced to the three-jet KK
cross section contain an extra factor of $\alpha_{{}_S}(Q)$
relative to the contribution of single on-shell $g^{\star}$
production. However, since virtual $g^{\star}$ exchange is
significant for dijet production, the many virtual $g^{\star}$
exchange diagrams leading to three jets in the final state --- for
which no external $g^{\star}$'s are produced on shell --- may also
have a significant effect. Although we do not calculate these
purely virtual exchange contributions here, we do note that they
would likely enhance our results.} arise from subprocesses where a
$g^{\star}$ is produced on-shell and subsequently decays into $q
\bar{q} $, \textit{e.g.}, via $q \bar{q}   \rightarrow g_n^{\star}
\rightarrow g_n^{\star}g \rightarrow  q \bar{q}  g$.
We concentrate on the production of the $g^{\star}$, postponing
the consideration of its subsequent decay for the meantime.  The
subprocesses satisfying five-momentum conservation for which a
$g^{\star}$ is produced on shell are:

\begin{figure}
\setlength{\abovecaptionskip}{0pt}
\centering{\includegraphics[bb=144 190 462 672]{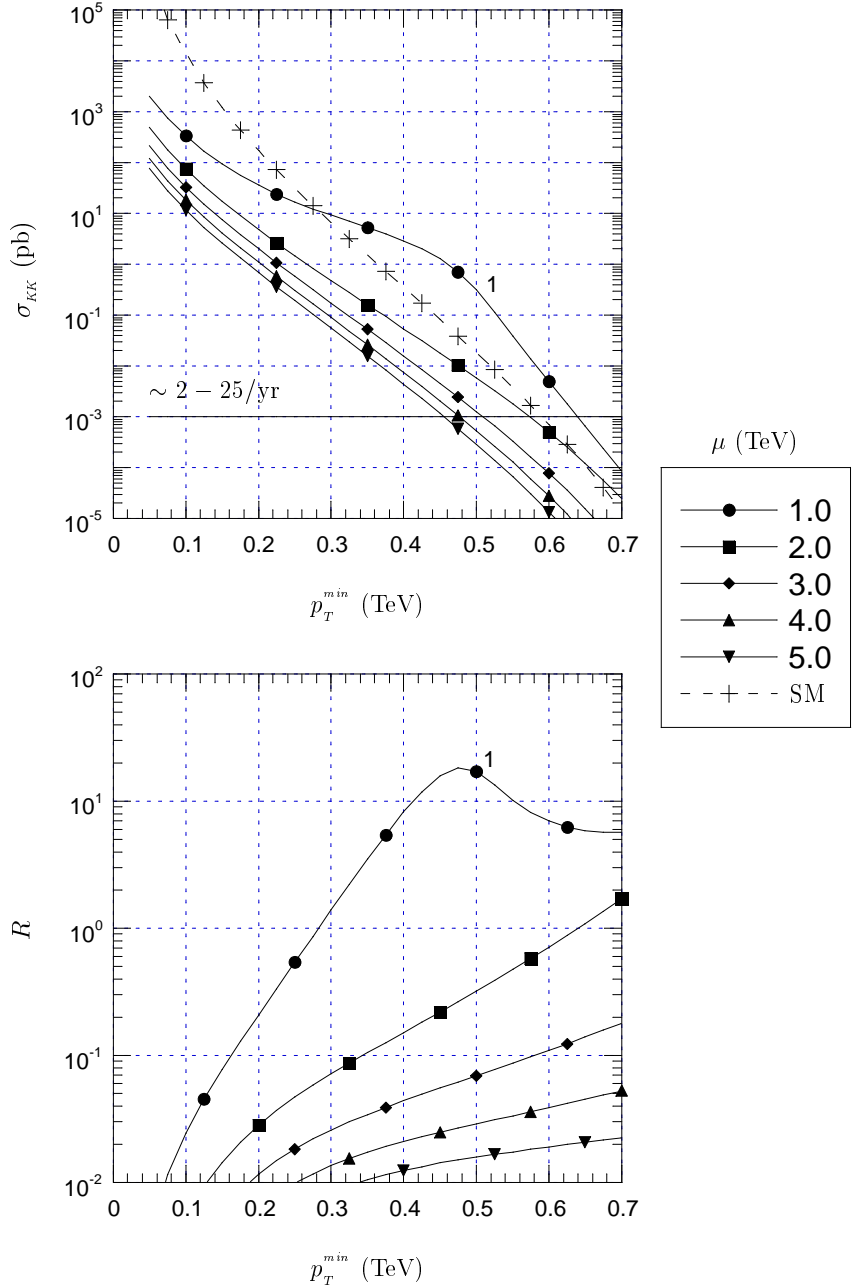}}
\vspace{10pt} \caption{The contributions of the virtual exchanges
of $g^{\star}$'s to the Tevatron dijet production cross section,
$\sigma_{{}_{\mathit{KK}}}  =  \sigma -
\sigma_{{}_{\mathit{SM}}}$, (top) and the ratio of the KK
contribution to the SM background, $R  = \sigma_{{}_{\mathit{KK}}}
/ \sigma_{{}_{\mathit{SM}}}$, (bottom) are illustrated as a
function of the minimum transverse momentum
$p_{{}_T}^{{}^{\mathit{min}}}$ for fixed values of the
compactification scale $\mu$. The solid horizontal line represents
$\sim 2$ ($25$) events/yr at the projected initial (final) Run 2
integrated luminosity.  Discernible bumps in regions for which
$p_{{}_T}^{{}^{\mathit{min}}} = k \mu / 2$ are indicated by the
corresponding value of $k \in \{1,2,\ldots\}$.}\label{fig:tgs2a}
\setlength{\abovecaptionskip}{0pt}
\end{figure}
\begin{figure}
\setlength{\abovecaptionskip}{0pt}
\centering{\includegraphics[bb=148 215 478 684]{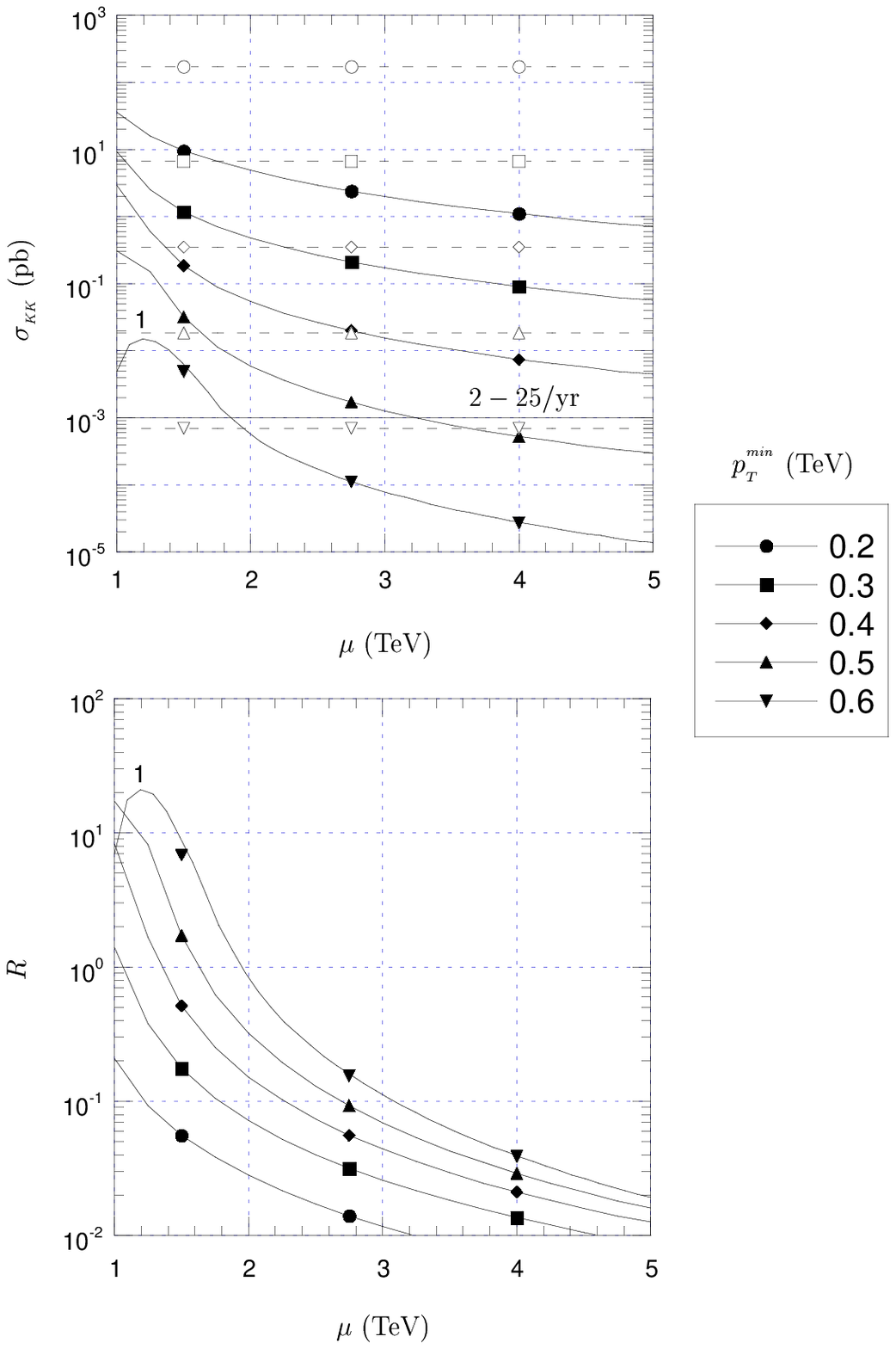}}
\vspace{10pt} \caption{Same as Fig.~\ref{fig:tgs2a}, but as a
function of the compactification scale $\mu$ for fixed values of
the minimum transverse momentum $p_{{}_T}^{{}^{\mathit{min}}}$.
The horizontal dashed lines represent the SM
background.}\label{fig:tgs2b} \setlength{\abovecaptionskip}{0pt}
\end{figure}

\begin{figure}
\centering{\includegraphics[bb=108 265.5 513 571.5]{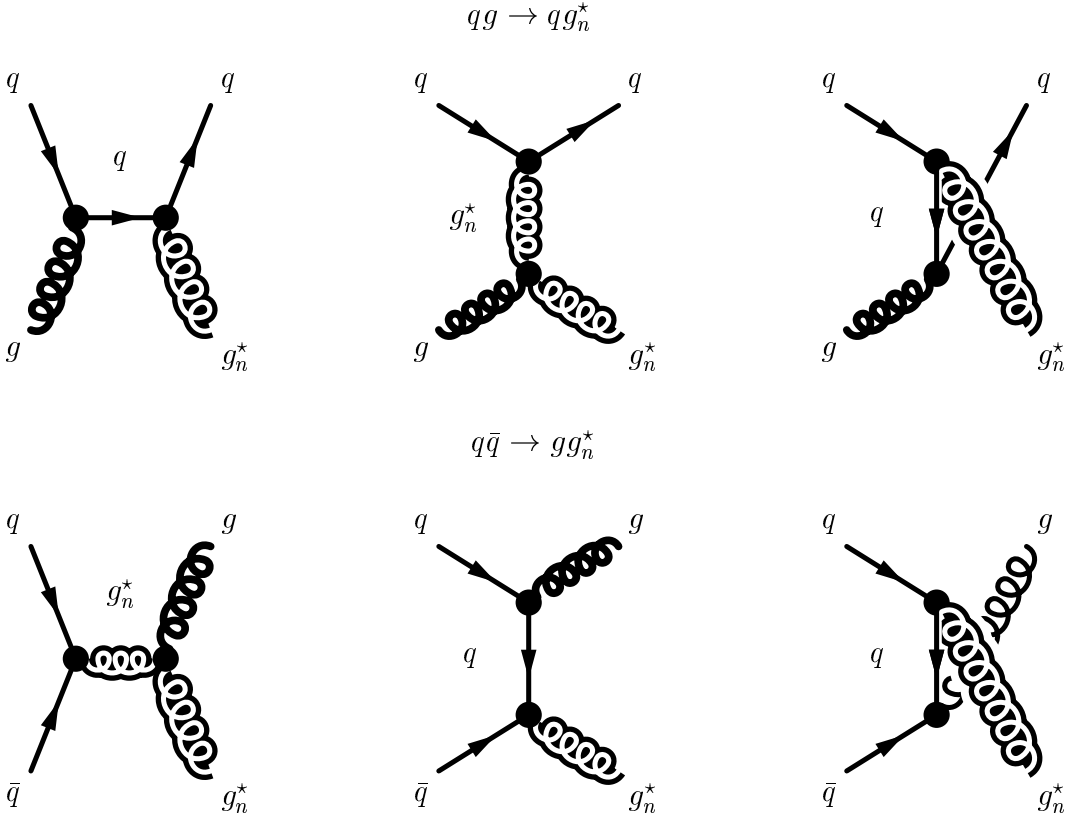}}
\vspace{0pt}\caption{Diagrams involving the production of a single
on-shell $g^{\star}$. The diagrams for $\bar{q} g \rightarrow
\bar{q} g_n^{\star}$ are obtained by replacing $q$ with $\bar{q}$
in the diagrams for $q g \rightarrow q g_n^{\star}$.}
\label{fig:3jets}
\end{figure}

\vspace{-11pt} \begin{eqnarray}
q \bar{q} \!\!&\rightarrow &\!\! g_n^{\star}g \nonumber \\
q g \!\!&\rightarrow &\!\! q g_n^{\star} \\
 \bar{q}  g \!\!&\rightarrow & \!\! \bar{q}  g_n^{\star} \, , \nonumber
\end{eqnarray}

\noindent where the mode $n$ of the external $g^{\star}$ is
necessarily identical to that of any virtual $g^{\star}$'s.
Therefore, there is no summation over modes in these propagators;
instead, the three-jet cross section involves a summation over the
possible modes ($n\geq 1$) of the external $g^{\star}$'s.  The
Feynman diagrams for these three KK subprocesses are illustrated
in Fig.~\ref{fig:3jets}.
The amplitude for $q \bar{q}  \rightarrow  g_n^{\star} g$ is

\vspace{-8pt} \begin{eqnarray} \label{eq:Mqqgg} \mathcal{M}(q
\bar{q}  \rightarrow  g_{n}^{\star} g) =\!\!\! & -& \!\!\!i 4 \pi
\alpha_{{}_S} (Q) \bar{v}_j (p_1) \mbox{\raisebox{-.6ex}{\huge
$[$}} T_{\mathit{ki}}^e T_{\mathit{jk}}^f
\mbox{\raisebox{-.6ex}{\huge $($}}\frac{V^t_{\rho \sigma}}{
\hat{t} }   +
\frac{V^s_{\rho \sigma}}{ \hat{s}'_n}\mbox{\raisebox{-.6ex}{\huge $)$}} \nonumber \\
\!\!\!&+& \!\!\!  T_{\mathit{ki}}^f T_{\mathit{jk}}^e
\mbox{\raisebox{-.6ex}{\huge $($}}\frac{V^u_{\rho \sigma}}{
\hat{u} }  -  \frac{V^s_{\rho \sigma}}{ \hat{s}
'_n}\mbox{\raisebox{-.6ex}{\huge $)$}}
\mbox{\raisebox{-.6ex}{\huge $]$}} u_i (p_2)
\epsilon_{e}^{\ast\rho} (k_1) \epsilon_{f}^{\ast\sigma} (k_2) \, ,
\end{eqnarray}

\noindent where the scale $Q$ is identified with the mass of the
$g^{\star}$, $ \hat{v} '_n$ represents (as before) subtraction of
$m_n^2$ from the Mandelstam variable $ \hat{v}   \in  \{ \hat{s} ,
\hat{t} , \hat{u} \}$ (\textit{i.e.}, $ \hat{v} '_n  =   \hat{v}
 -  m_n^2$), and the $V^{v}_{\rho \sigma}$ tensors are given
by

\pagebreak[2] \vspace{-8pt} \begin{eqnarray} V^s_{\rho \sigma}
\!\!\! &=& \!\!\! \sqrt{2}\gamma^{\mu}
\mbox{\raisebox{-.6ex}{\huge $[$}}(k_2  +  2k_1)_{\sigma}
g_{\mu\rho}  +
(-k_1  +  k_2)_{\mu} g_{\rho\sigma}  -  (2k_2+k_1)_{\rho} g_{\sigma\mu}\mbox{\raisebox{-.6ex}{\huge $]$}} \\
V^t_{\rho \sigma} \!\!\! &=& \!\!\! \sqrt{2}\gamma_{\rho} (\not \! p_1  -  \not \! k_1) \gamma_{\sigma}\\
V^u_{\rho \sigma} \!\!\! &=& \!\!\! \sqrt{2}\gamma_{\sigma} (\not
\! k_1 -  \not \! p_2) \gamma_{\rho} \, .
\end{eqnarray}

\noindent After summing over final states and averaging over
initial states, the resulting ampli\-tude-squared
is\addtocounter{footnote}{-1}\footnote{We employ
\textit{FORM}~\cite{FORM}, a symbolic manipulation program, in the
evaluation of the amplitudes-squared for single and double
on-shell $g^{\star}$ production.}

\vspace{-8pt} \begin{equation}
\mbox{\raisebox{-.5ex}{\Large$\bar{\Sigma}$}}\! \mid \!
\mathcal{M}(q \bar{q} \rightarrow g_n^{\star} g)\! \mid ^2 \,
  =  \frac{8}{27}\pi^2 \alpha_S^2(Q)\mbox{\raisebox{-.6ex}{\huge
$[$}}\mbox{\raisebox{-.6ex}{\huge $($}}\frac{m_n^4}{ \hat{s} '
{}_n^2}  +  \frac{m_n^2}{ \hat{s} '_n}\mbox{\raisebox{-.6ex}{\huge
$)$}} \mbox{\raisebox{-.6ex}{\huge $($}}8\frac{ \hat{s} ' {}_n^2}{
\hat{t}   \hat{u} }  -  18\mbox{\raisebox{-.6ex}{\huge $)$}}\!
  -  17  +  4\frac{ \hat{s} ' {}_n^2}{ \hat{t}   \hat{u} }  +  18\frac{ \hat{t}   \hat{u} }{ \hat{s} ' {}_n^2}\mbox{\raisebox{-.6ex}{\huge
$]$}} \, ,
\end{equation}

\noindent which is related to the amplitude-squared for $qg
\rightarrow  qg_n^{\star}$ via crossing symmetry:

\vspace{-3pt} \begin{equation}
\mbox{\raisebox{-.5ex}{\Large$\bar{\Sigma}$}} \! \mid \!
\mathcal{M}(qg \rightarrow qg_n^{\star})\! \mid ^2 \, =
\frac{1}{9}\pi^2 \alpha_S^2(Q)\mbox{\raisebox{-.6ex}{\huge
$[$}}\mbox{\raisebox{-.6ex}{\huge $($}}\frac{m_n^4}{ \hat{u} '
{}_n^2}  +  \frac{m_n^2}{ \hat{u} '_n}\mbox{\raisebox{-.6ex}{\huge
$)$}} \mbox{\raisebox{-.6ex}{\huge $($}}18  -  8\frac{ \hat{u} '
{}_n^2}{ \hat{s}   \hat{t} }\mbox{\raisebox{-.6ex}{\huge $)$}}\!
  +  17  -  4\frac{ \hat{u} ' {}_n^2}{ \hat{s}   \hat{t} }  +  18\frac{ \hat{s}   \hat{t} }{ \hat{u} ' {}_n^2}\mbox{\raisebox{-.6ex}{\huge
$]$}} \, .
\end{equation}

\noindent The amplitude-squared for $ \bar{q}  g  \rightarrow
\bar{q}  g_n^{\star}$ is in turn identical to that of $qg
\rightarrow qg_n^{\star}$ by time-reversal invariance.  Upon
integration over $ \hat{t} $, the single $g^{\star}$ on-shell
production cross sections assume the form

\vspace{-8pt} \begin{eqnarray} \sigma_{{}_{\mathit{KK}}}(p p
\rightarrow  g^{\star}  + \mathrm{jet})  =  \frac{1}{2\pi}
\sum_{j}\sum_{g_n^{\star}} & \!\!\!\! & \!\!\!\!
\,\,\,\int_{\mbox{\raisebox{-1.5ex}{\scriptsize{$\!\!\!\!\!\!\!\!\!\!
m_n^2/ s $}}}}^{\mbox{\raisebox{.9ex}{\scriptsize{$\!\!\! 1$}}}}
dx_{{}_A}
\,\,\,\,\int_{\mbox{\raisebox{-1.5ex}{\scriptsize{$\!\!\!\!\!\!\!\!\!\!\!\!\!\!
m_n^2/ s \, x_{{}_A}
$}}}}^{\mbox{\raisebox{.9ex}{\scriptsize{$\!\!\!
1$}}}}dx_{{}_B} f_{a/{}_A}(x_{{}_A},Q) \nonumber \\
& \!\!\!\! & \!\!\!\! f_{b/{}_B}(x_{{}_B},Q)
\int_{\mbox{\raisebox{-1.3ex}{\scriptsize{$\!\!\!\!\!\!
-1$}}}}^{\mbox{\raisebox{.9ex}{\scriptsize{$\!\!\! 1$}}}}d z \,
\mbox{\raisebox{-.5ex}{\Large$\bar{\Sigma}$}}\! \mid \! M_{jn} \!
\mid ^2 \, \frac{ \hat{s} '_n}{ \hat{s} ^2} \, ,
\end{eqnarray}

\noindent where the first summation runs over all possible
subprocesses $j$ producing a single $g^{\star}$ on-shell, and the
second summation is over all $g_n^{\star}$'s that can be produced
for subprocess $j$ in light of the given $pp$ collider energy
$\sqrt{s}$.\footnote{Note that the scale $Q = m_n$ for the $n
> 1$ modes exceeds the compactification scale $\mu$. When $Q >
\mu$, the running of $\alpha_{{}_S} (Q)$ transforms from a
logarithmic to a power law behavior~\cite{unify}.  This has the
effect of reducing the contributions of the higher order modes to
the total multijet cross sections~\cite{topview}, but only
slightly at LHC energies since only a few KK modes can be produced
on-shell.}  Observe that $M_{jn}(m_n) = M_{j1}(nm_1)$ so that
$\sum_{n=1}^{n_{{}_{\mathit{max}}}} M_{jn}(m_n) =
\sum_{n=1}^{n_{{}_{\mathit{max}}}} M_{j1}(nm_1)$. We are now
prepared to account for the decay of the $g_n^{\star}$ into $q
\bar{q} $ pairs. Working in the narrow width approximation, we
integrate over the dimensionless solid angle $d\Omega_4/4\pi$ to
obtain the total single on-shell $g^{\star}$ cross section (prior
to cuts):

\vspace{-3pt} \begin{equation} \sigma_{{}_{\mathit{KK}}}(pp
\rightarrow  \mathrm{jet}  + g^{\star}\rightarrow 3 \,
\mathrm{jets})  = \int
\frac{d\Omega_4}{4\pi}{\sigma_{{}_{\mathit{KK}}} (pp  \rightarrow
g^{\star}+\mathrm{jet})} \, .
\end{equation}

\noindent The various cuts are performed by defining the two
4-momenta of the decaying particles in their center of mass frame
in terms of $\Omega_4$ (each decaying particle has momentum
$m_n/2$) and boosting the two 4-momenta to the lab frame. In
addition to the $g^{\star}$ cross section, we calculate the SM
three-jet background following the outline of Ref.~\cite{SM3jet}.

\begin{figure}
\setlength{\abovecaptionskip}{0pt}
\centering{\includegraphics[bb=144 195 471 672]{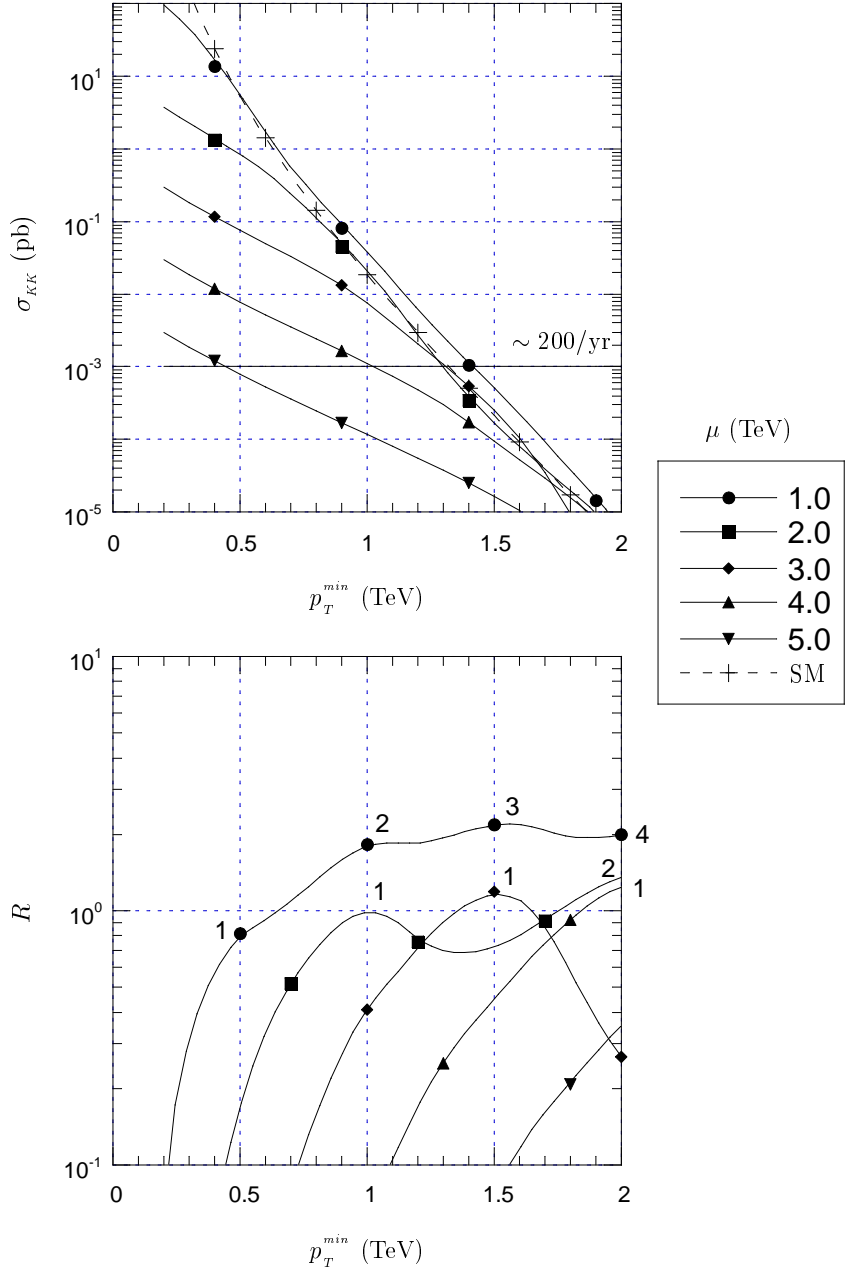}}
\vspace{10pt} \caption{The contributions of the single-on shell
production of $g^{\star}$'s to the three-jet cross section at the
LHC, $\sigma_{{}_{\mathit{KK}}}  =  \sigma -
\sigma_{{}_{\mathit{SM}}}$, (top) and the ratio of the KK
contribution to the SM background, $R  = \sigma_{{}_{\mathit{KK}}}
/ \sigma_{{}_{\mathit{SM}}}$, (bottom) are illustrated as a
function of the minimum transverse momentum
$p_{{}_T}^{{}^{\mathit{min}}}$ for fixed values of the
compactification scale $\mu$. The solid horizontal line represents
$\sim 200$ events/yr at the projected integrated luminosity.
Discernible bumps in regions for which
$p_{{}_T}^{{}^{\mathit{min}}} = k \mu / 2$ are indicated by the
corresponding value of $k \in \{1,2,\ldots\}$. }\label{fig:gs3a}
\setlength{\abovecaptionskip}{0pt}
\end{figure}

\begin{figure}
\setlength{\abovecaptionskip}{0pt}
\centering{\includegraphics[bb=153 238 458 688]{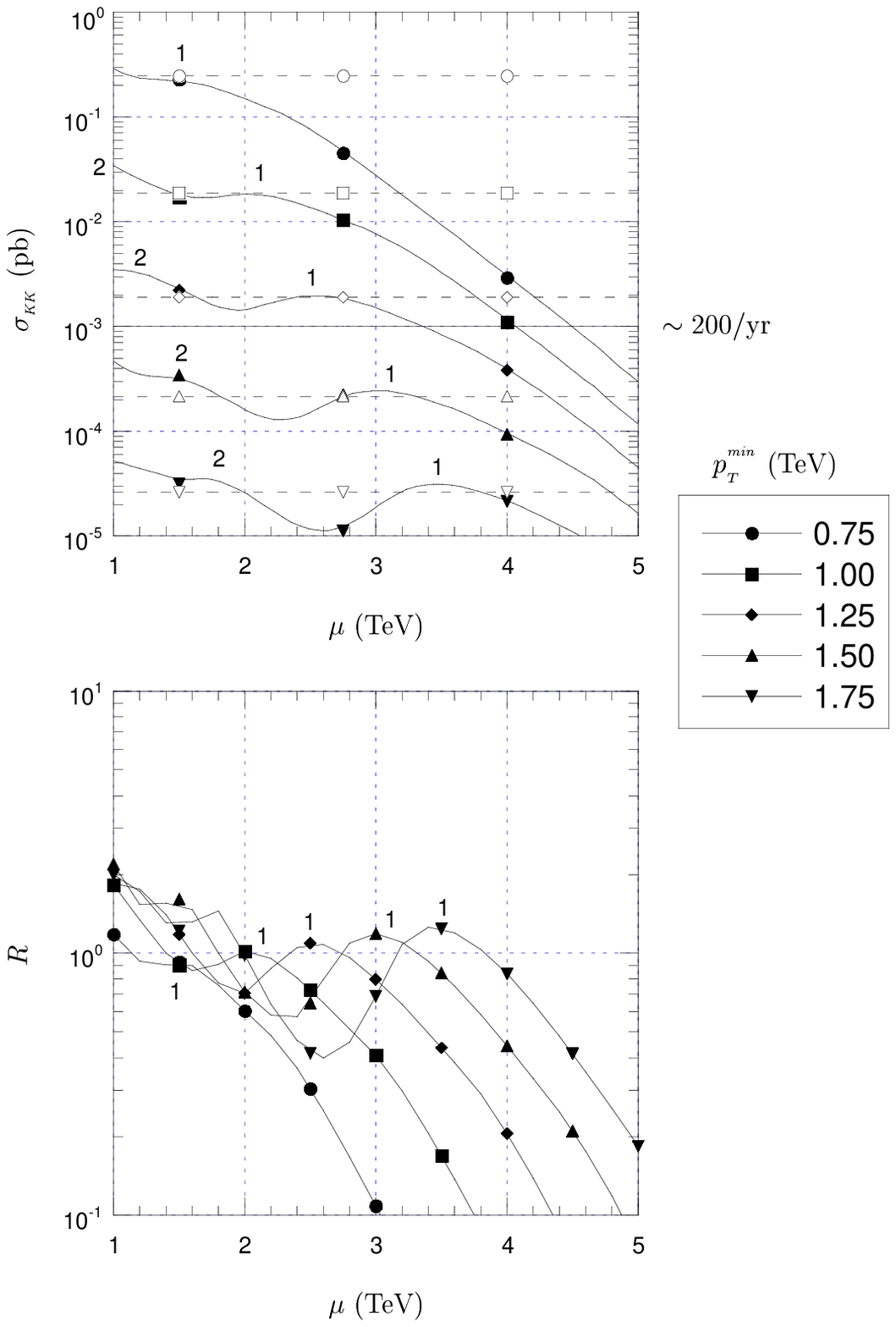}}
\vspace{10pt} \caption{Same as Fig.~\ref{fig:gs3a}, but as a
function of the compactification scale $\mu$ for fixed values of
the minimum transverse momentum $p_{{}_T}^{{}^{\mathit{min}}}$.
The horizontal dashed lines represent the SM
background.}\label{fig:gs3b} \setlength{\abovecaptionskip}{0pt}
\end{figure}

\begin{figure}
\setlength{\abovecaptionskip}{0pt}
\centering{\includegraphics[bb=123 460 498 698]{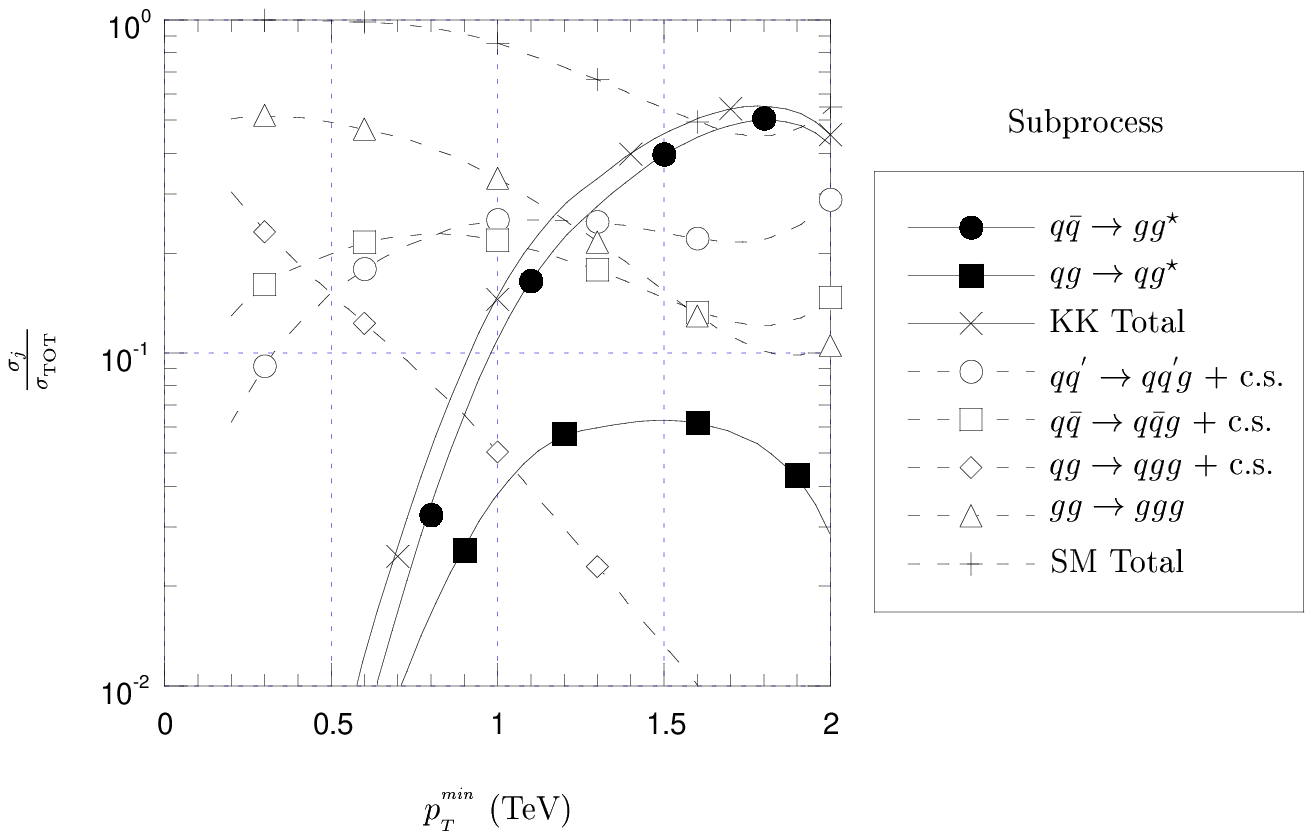}}
\vspace{10pt} \caption{The partial contributions to the total
three-jet cross section are shown as a function of
$p_{{}_T}^{{}^{\mathit{min}}}$, for $\mu = 3.5$ TeV. Here, c.s.
represents all subprocesses that are related by crossing
symmetry.}\label{fig:sm3p} \setlength{\abovecaptionskip}{0pt}
\end{figure}

\begin{figure}
\setlength{\abovecaptionskip}{0pt}
\centering{\includegraphics[bb=121 460 463 696]{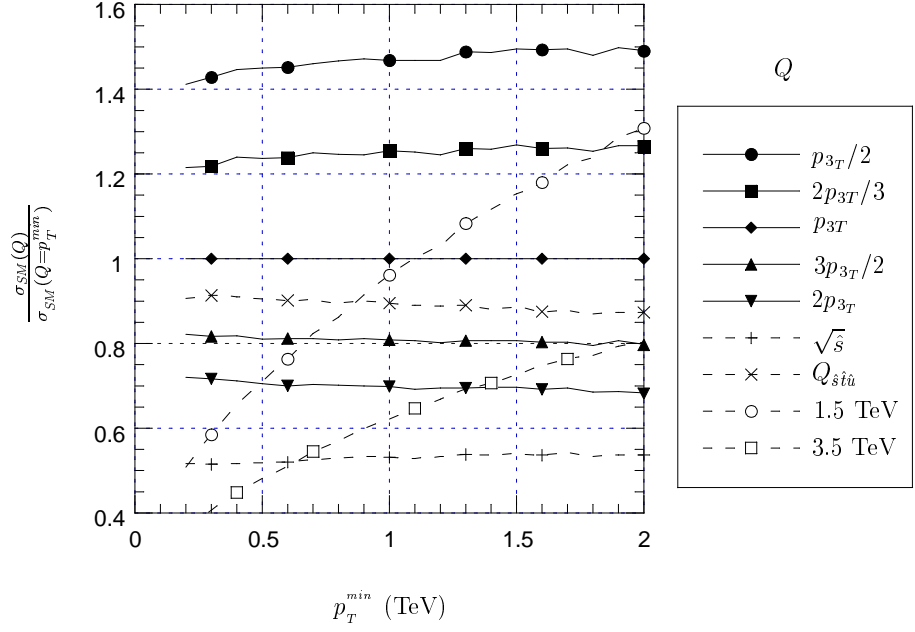}}
\vspace{10pt} \caption{The effect that variation of the choice of
$Q$ has on the SM three-jet background is shown as a function of
the minimum transverse momentum, $p_{{}_T}^{{}^{\mathit{min}}}$.
Here, $p_{3T}$ is the transverse momentum of one of the jets,
$Q_{\hat{s}\hat{t}\hat{u}} =
\sqrt{\frac{\hat{s}\hat{t}\hat{u}}{\hat{s}^2  +  \hat{t}^2  +
\hat{u}^2}}$, and values in TeV (\textit{e.g.}, $3.5$ TeV)
correspond to the choice of (constant) $Q$ equal to a
compactification scale at that particular scale.} \label{fig:sm3q}
\setlength{\abovecaptionskip}{0pt}
\end{figure}

In addition to the cuts applied for dijet production, for three or
four jets, we constrain final states to be separated by a cone of
radius $R = \sqrt{(\Delta\phi) + (\Delta\eta)} = 0.4$, where
$\phi$ is the azimuthal angle and $\eta$ is the pseudorapidity,
which is related to the polar angle $\theta$ via $\eta = - \ln
\tan (\theta/2)$. The single on-shell $g^{\star}$ production cross
sections, along with the SM background, are plotted in
Fig.'s~\ref{fig:gs3a}--\ref{fig:gs3b} for $1$ TeV $\leq \mu \leq
5$ TeV and $p_{{}_T}^{{}^{\mathit{min}}}  \leq 2$ TeV.
 High $p_{{}_T}$ cuts have a similar effect to that
described for dijet production except that the
$p_{{}_T}^{{}^{\mathit{min}}} = k \mu / 2$ disturbances are much
larger than the dijet case, which should be expected since the
$g^{\star}$ is produced on-shell in the three-jet case considered
here.  Such discernible disturbances are indicated by the
corresponding values of $k \in \{1,2, \ldots\}$. Again we
terminate the $p_{{}_T}$ cuts when the number of anticipated
events is quite scarce ($\sim 1/$yr). Although it is not as
extreme as in the dijet case, the single on-shell $g^{\star}$
results also exceed the SM background for very high
$p_{{}_T}^{{}^{\mathit{min}}}$. The partial contributions of the
various subprocesses to the $g^{\star}$ (for a representative
value of $\mu = 3.5$ TeV) and SM cross sections are shown in
Fig.~\ref{fig:sm3p}.
The $qg\rightarrow qg^{\star}$ subprocess dominates over the range
of interest, and $q\bar{q}\rightarrow gg^{\star}$ only contributes
to the KK dijet cross section significantly for low $p_{{}_T}$.
The effect of varying $Q$ in the SM for three jets resembles the
effect for two jets to a large degree (Fig.~\ref{fig:sm3q}).

We point out that our calculation of the background for these
three-jet final states is somewhat of an overestimate.  In our
signal, two of the jets come from the decay of an on-shell
$g^{\star}$.  If we impose the condition that two of the jets
cluster around the $g^{\star}$ mass for the SM background, the
background to the signal ratio will be less.  We did not impose
that since we are not certain whether that will be possible to
implement experimentally in the actual detection of the jets.  If
that is experimentally feasible, the background to our signal
ratio will be less.

\vspace{0.5cm}

\noindent {\bf 5.  Double On-Shell $g^{\star}$ Production}

\vspace{0.2cm}

\begin{figure}
\centering{\includegraphics[bb=108 67.5 513 675]{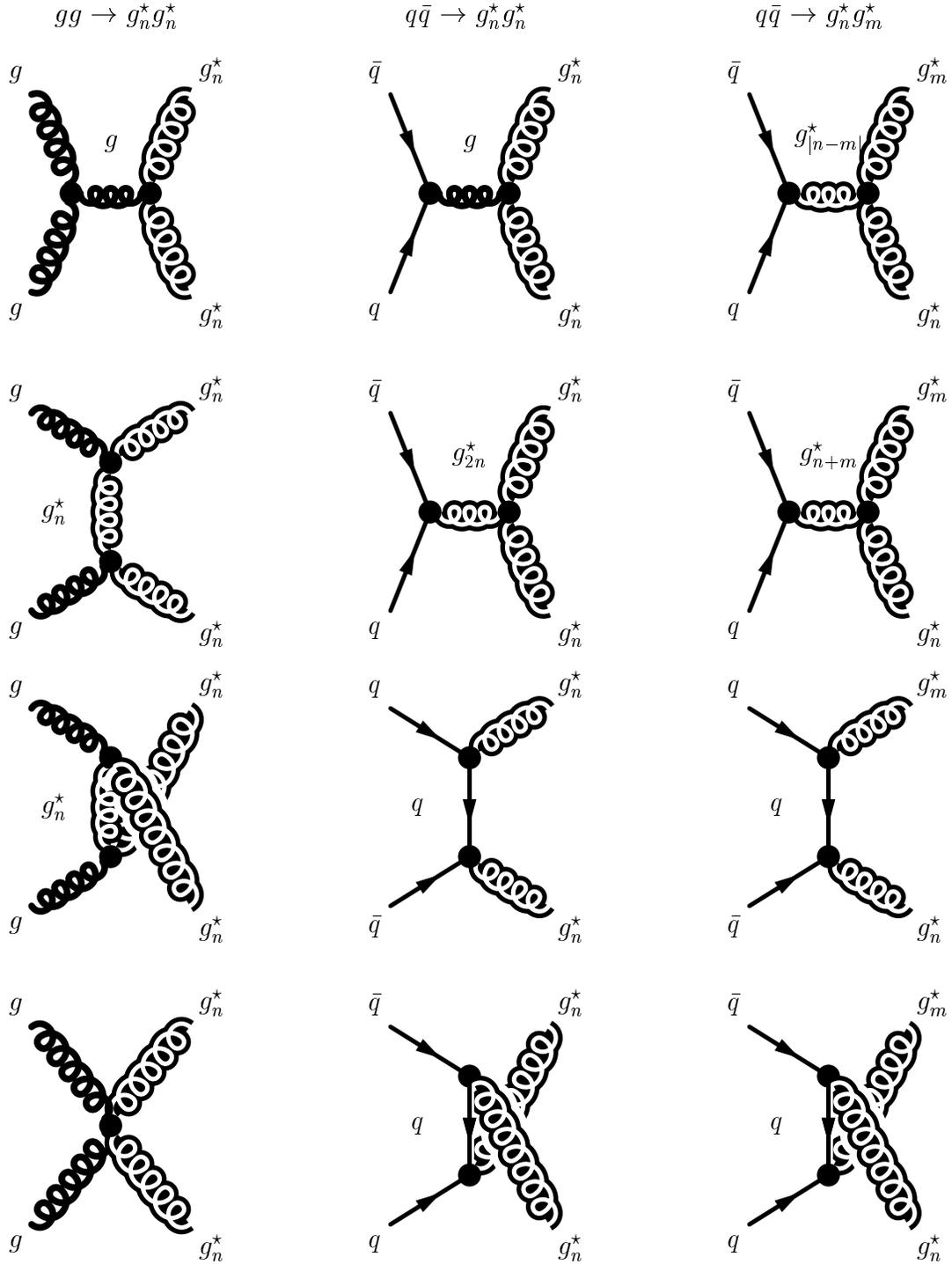}}
\vspace{-35pt} \caption{Diagrams involving the production of two
on-shell $g^{\star}$'s. The modes $n$ and $m$ are distinct ($n
\neq m$).} \label{fig:4jets}
\end{figure}

\noindent Double on-shell $g^{\star}$ production is analogous to
single on-shell $g^{\star}$ production, except that in this case
the predominant KK subprocesses involve the production of two
on-shell $g^{\star}$'s which subsequently decay into $q \bar{q} $
pairs, \textit{e.g.}, $q \bar{q}   \rightarrow  g\rightarrow
g_n^{\star}g_n^{\star}  \rightarrow  q \bar{q}  q \bar{q} $. Also,
the single on-shell $g^{\star}$ case did not involve the
$g_n^{\star}$-$g_m^{\star}$-$g_{ \ell }^{\star}$ nor the
$g$-$g$-$g_n^{\star}$-$g_n^{\star}$ vertices, which are now part
of the picture. Focusing on the production of the $g^{\star}$'s
for the present and applying five-momentum conservation, the
subprocesses for which two $g^{\star}$'s are produced on shell
are:

\vspace{-11pt} \begin{eqnarray}
q \bar{q} \!\!\!  &\rightarrow &\!\!\! g_n^{\star}g_n^{\star} \nonumber \\
q \bar{q} \!\!\! &\rightarrow &\!\!\! g_n^{\star}g_m^{\star} \\
g g \!\!\! &\rightarrow &\!\!\! g_n^{\star}g_n^{\star} \, ,
\nonumber
\end{eqnarray}

\noindent where the two external $g^{\star}$'s are necessarily in
the same mode $n$ for initial gluons, but not for initial quarks.
The Feynman diagrams for these three KK subprocesses are
illustrated in Fig.~\ref{fig:4jets}.  The diagrams for $q \bar{q}
\rightarrow  g_n^{\star}g_n^{\star}$ are the same as for $q
\bar{q}   \rightarrow  g_n^{\star}g$ except that the $ \hat{s}
$-channel diagram can have either a virtual $g$ or a virtual
$g_{2n}^{\star}$ propagator.  Thus, the amplitude for this process
is the same as that given by Eq.~(\ref{eq:Mqqgg}) with the
$g_n^{\star}$ propagator replaced by $g$ and $g_{2n}^{\star}$
propagators, where the coefficient of the $ \hat{s} $-channel
amplitude is reduced by $1/\sqrt{2}$ for the $g$ case.  Likewise,
the subprocess $q \bar{q}  \rightarrow g_n^{\star}g_m^{\star}$ is
simply $q \bar{q}  \rightarrow g_n^{\star}g_n^{\star}$ with the
$s$-channel altered for the possible propagators and the mass of
either external line altered by a factor of $m/n$. The amplitude
for $gg  \rightarrow g_n^{\star}g_n^{\star}$ is

\vspace{-8pt} \begin{eqnarray} \mathcal{M}(g g  \rightarrow
g_{n}^{\star} g_{n}^{\star}) =& \!\!\! -& \!\!\!i 4 \pi
\alpha_{{}_S} (Q) \mbox{\raisebox{-.6ex}{\huge $($}}
f^{\mathit{abc}}f^{\mathit{cef}}\frac{V^s_{\alpha\beta\rho\sigma}}{
\hat{s} } \, + \,
f^{\mathit{bec}}f^{\mathit{acf}}\frac{V^t_{\alpha\beta\rho\sigma}}{
\hat{t} '_n}
 \nonumber \\ & \!\!\!   +  & \!\!\!
f^{\mathit{bfc}}f^{\mathit{ace}}\frac{V^u_{\alpha\beta\rho\sigma}}{
\hat{u} '_n}
  +  V^{4\mathit{abef}}_{\alpha\beta\rho\sigma}\mbox{\raisebox{-.6ex}{\huge $)$}}
   \epsilon_{a}^{\alpha}(p_1)\epsilon_{b}^{\beta}
(p_2)\epsilon_{e}^{\ast\rho} (k_1) \epsilon_{f}^{\ast\sigma} (k_2)
\, ,
\end{eqnarray}

\noindent
where

\begin{eqnarray} V^s_{\alpha\beta\rho\sigma}& \!\!\!
=& \!\!\! \mbox{\raisebox{-.6ex}{\huge $[$}}(-p_1  + p_2)_{\mu}
g_{\alpha\beta}  +  (2p_1  +  p_2)_{\alpha} g_{\beta\mu}
  -  (p_1+2p_2)_{\beta} g_{\mu\alpha}\mbox{\raisebox{-.6ex}{\huge $]$}} \nonumber \\
& \!\!\!  & \!\!\! \cdot \mbox{\raisebox{-.6ex}{\huge $[$}}(2k_1
\, + \, k_2)_{\sigma} g_{\nu\rho}  +
(-k_1  +  k_2)_{\nu} g_{\rho\sigma}  -  (k_1+2k_2)_{\rho} g_{\sigma\nu}\mbox{\raisebox{-.6ex}{\huge $]$}}g^{\mu\nu}\\
V^t_{\alpha\beta\rho\sigma}& \!\!\! =& \!\!\!
\mbox{\raisebox{-.6ex}{\huge $[$}}(p_1  +  k_1)_{\mu}
g_{\beta\rho}  +  (p_1  -  2k_1)_{\beta} g_{\rho\mu}  +
(-2p_1+k_1)_{\rho} g_{\mu\beta}\mbox{\raisebox{-.6ex}{\huge $]$}} \nonumber \\
& \!\!\!  & \!\!\! \cdot \mbox{\raisebox{-.6ex}{\huge $[$}}(2p_2
\, - \, k_2)_{\sigma} g_{\alpha\nu}  + (-p_2  +  2k_2)_{\alpha}
g_{\nu\sigma}  -  (p_2+k_2)_{\nu}
g_{\alpha\sigma}\mbox{\raisebox{-.6ex}{\huge $]$}}g^{\mu\nu}\\
V^u_{\alpha\beta\rho\sigma}& \!\!\! =& \!\!\!
\mbox{\raisebox{-.6ex}{\huge $[$}}(p_1  +  k_2)_{\mu}
g_{\beta\sigma}  +  (p_1  -  2k_2)_{\beta} g_{\sigma\mu}
  +  (-2p_1+k_2)_{\sigma} g_{\mu\beta}\mbox{\raisebox{-.6ex}{\huge $]$}} \nonumber \\
& \!\!\!  & \!\!\! \cdot \mbox{\raisebox{-.6ex}{\huge $[$}}(2p_2
\, - \, k_1)_{\rho} g_{\alpha\nu}  + (-p_2  +  2k_1)_{\alpha}
g_{\nu\rho}  -  (p_2+k_1)_{\nu}
g_{\alpha\rho}\mbox{\raisebox{-.6ex}{\huge $]$}}g^{\mu\nu}
\end{eqnarray}

\begin{eqnarray}
V^{4\mathit{abef}}_{\alpha\beta\rho\sigma}&\!\!\! =& \!\!\!
f_{\mathit{abc}}f_{\mathit{efc}}(g_{\alpha\rho}g_{\beta\sigma}  -
g_{\alpha\sigma}g_{\beta\rho})  +
f_{\mathit{aec}}f_{\mathit{fbc}}(g_{\alpha\sigma}g_{\beta\rho}  -
g_{\alpha\beta}g_{\sigma\rho}) \nonumber \\
& \!\!\! & \!\!\!   +
f_{\mathit{afc}}f_{\mathit{bec}}(g_{\alpha\beta}g_{\sigma\rho} \,
- \, g_{\alpha\rho}g_{\beta\sigma}) \, .
\end{eqnarray}

\noindent The amplitudes-squared for these subprocesses, summed
over final states and averaged over initial states, are obtained
to be

\vspace{-8pt} \begin{eqnarray}
\mbox{\raisebox{-.5ex}{\Large$\bar{\Sigma}$}}\! \mid \!
\mathcal{M}(q \bar{q}   \rightarrow g_n^{\star} g_n^{\star}) \!
\mid ^2 \, =& \!\!\! & \!\!\!\!\!\!\!\!\!\frac{8}{27}\pi^2
\alpha_S^2(Q)\mbox{\raisebox{-.6ex}{\huge $[$}}648 m_n^6\frac{1}{
\tilde{s}_n  \hat{t}\hat{u}}  -  738 m_n^4
\mbox{\raisebox{-.6ex}{\huge $($}}\frac{1}{\tilde{s}_n \hat{s}}
  -  27\frac{1}{ \hat{s}^2 }  +  164\frac{1}{ \hat{t} \hat{u}} \nonumber \\
& \!\!\! -& \!\!\!  16\frac{1}{ \hat{t} ^2}  -   16\frac{1}{ \hat{u} ^2} - \,  27\frac{1}{ \tilde{s}_n  ^2} \mbox{\raisebox{-.6ex}{\huge $)$}} \nonumber\\
 & \!\!\! +& \!\!\! 9 m_n^2 \mbox{\raisebox{-.6ex}{\huge $($}} 32\frac{ \hat{s} }{ \hat{t}   \hat{u} }  -  144\frac{1}{ \hat{s} }\mbox{\raisebox{-.6ex}{\huge $)$}}\!  -  68
  +  16\frac{ \hat{s} ^2}{ \hat{t}   \hat{u} }  +  18\frac{ \hat{t}   \hat{u} }{ \hat{s} \tilde{s}_n } \nonumber \\
& \!\!\! +& \!\!\! 27\frac{ \hat{t}   \hat{u} }{ \hat{s} ^2}  + 27
\frac{ \hat{t}   \hat{u} }{ \tilde{s}_n ^2 }
\mbox{\raisebox{-.6ex}{\huge $]$}}
\end{eqnarray}

\begin{eqnarray}
\mbox{\raisebox{-.5ex}{\Large$\bar{\Sigma}$}}\! \mid \!
\mathcal{M}(q \bar{q}   \rightarrow  g_n^{\star} g_m^{\star}) \!
\mid ^2 \, =& \!\!\!& \!\!\!\!\!\!\!\!\!
\frac{8}{27}\pi^2 \alpha_S^2(Q)\mbox{\raisebox{-.6ex}{\huge $[$}}-14 \frac{\hat{s} \hat{t}^2}{\hat{u}}  +  2 \frac{\hat{t}^3}{\hat{u}}  -  20 \hat{t} \hat{u} \nonumber \\
&\!\!\! +&\!\!\! \mbox{\raisebox{-.6ex}{\huge $($}}\!\!\!-8
\frac{\hat{t}^2}{\hat{u}^2} m_m^2 m_n^2  +  30 \hat{t} m_m^2
+14 \frac{\hat{t}}{\hat{u}} m_m^4  -  25 \frac{\hat{t}}{\hat{u}} m_m^2 m_n^2 \nonumber \\
&\!\!\! -&\!\!\! 44 m_m^4
 -  24 m_m^2 m_n^2  -  16 \frac{m_m^4 m_n^2}{\hat{u}} \nonumber \\
&\!\!\!+&\!\!\! 32 \frac{m_m^4 m_n^4}{\hat{u}^{2}}  +  8 \frac{m_m^6 m_n^2}{\hat{t} \hat{u}}  +  8 \frac{m_m^4 m_n^4}{\hat{t} \hat{u}} \nonumber \\
&\!\!\!+&\!\!\! m \leftrightarrow n\mbox{\raisebox{-.6ex}{\huge $)$}}\! + t \leftrightarrow u\mbox{\raisebox{-.6ex}{\huge $]$}} \frac{1}{\hat{s}-(m_m+m_n)^2}\frac{1}{\hat{s}-(m_m-m_n)^2}\\
\mbox{\raisebox{-.5ex}{\Large$\bar{\Sigma}$}}\! \mid \!
\mathcal{M}(g g  \rightarrow  g_n^{\star} g_n^{\star})\!\mid ^2 \,
=& \!\!\!& \!\!\!\!\!\!\!\!\! \frac{9}{4}\pi^2
\alpha_S^2(Q)\mbox{\raisebox{-.6ex}{\huge $($}}\frac{s^2}{ \hat{t}
'_n  \hat{u} '_n}  - 1\mbox{\raisebox{-.6ex}{\huge $)$}}
\nonumber \\
&\!\!\! \cdot &\!\!\!\!\!\! \mbox{\raisebox{-.6ex}{\huge
$($}}6\frac{m_n^4}{ \hat{t} '_n \hat{u} '_n}
  -  6\frac{m_n^2}{ \hat{s} }  +  2\frac{ \hat{s} ^2}{ \hat{t} '_n  \hat{u} '_n}  +
\frac{ \hat{t} '_n  \hat{u} '_n}{ \hat{s} ^2}  -
4\mbox{\raisebox{-.6ex}{\huge $)$}} \, ,
\end{eqnarray}

\noindent where $\tilde{s}_n  \equiv  \hat{s}  -  4 m_n^2$.  (In
our notation, the replacements indicated by $t \leftrightarrow u$
do not affect the two terms that involve neither $t$ nor $u$.)

We point out that in our results for the matrix element squares,
as given in Eqs. $(29-31)$, there are no terms that grow with
energy, and the matrix elements for these subprocesses are
tree-unitary.  This is not true for the individual diagrams for
the subprocesses:  There are delicate cancellations between the
diagrams for each subprocess.  These cancellations occur only
because of the relations among the couplings as dictated by the
compactification of the five-dimensional KK theory to four
dimensions, and also due to the special relations for the masses
of the various KK states.  For example, in the process $q \bar{q}
\rightarrow g_n^{\star} g_n^{\star}$, the presence of the
$g_{2n}^{\star}$ exchange is crucial with its mass $2 n \mu$ and
its coupling as dictated by the KK Yang-Mills theory.  This is a
new example of tree-unitarity for a class of massive vector boson
theories other than the known spontaneously broken gauge
theories~\cite{unitary}.

These subprocess $j$ amplitudes-squared combine to give the total
KK cross section for $g_n^{\star}$'s produced on-shell as

\vspace{-8pt} \begin{eqnarray} \sigma_{{}_{\mathit{KK}}}(p p
\rightarrow  g^{\star}g^{\star})
 =  \frac{1}{4\pi} & \!\!\!\!\! & \!\!\!\!\!\!\!
\sum_{j}\sum_{g^{\star} \mathrm{pairs}} \,\,\,\,
\int_{\mbox{\raisebox{-1.5ex}{\scriptsize{$\!\!\!\!\!\!\!\!
\rho_{\mathit{mn}}$}}}}^{\mbox{\raisebox{.9ex}{\scriptsize{$\!\!\!
1$}}}}dx_{{}_A}
\int_{\mbox{\raisebox{-1.5ex}{\scriptsize{$\!\!\!\!\!\!\!\!\!\!
\rho_{\mathit{mn}}/x_{{}_A}$}}}}^{\mbox{\raisebox{.9ex}{\scriptsize{$\!\!\!
1$}}}}\!\!\!\!dx_{{}_B}
f_{a/{}_A}(x_{{}_A},Q) f_{b/{}_B}(x_{{}_B},Q) \nonumber \\
& \!\! & \!\!
\int_{\mbox{\raisebox{-1.3ex}{\scriptsize{$\!\!\!\!\!\!
-1$}}}}^{\mbox{\raisebox{.9ex}{\scriptsize{$\!\!\! 1$}}}} d z \,
\mbox{\raisebox{-.5ex}{\Large$\bar{\Sigma}$}}\! \mid \! M_j \!
\mid ^2 \, \frac{1}{ \hat{s} }\sqrt{1-\frac{(m_m+m_n)^2}{ \hat{s}
}} \, ,
\end{eqnarray}

\noindent where $\rho_{\mathit{mn}}  =  (m_m  +  m_n)^2 / s$ and
the second summation runs over all $g_n^{\star},g_m^{\star}$ pairs
that can be produced for energy $\sqrt{s}$. Again, we apply the
narrow width approximation to account for the decay of the
$g^{\star}$'s into $q \bar{q} $ pairs:

\vspace{-3pt} \begin{equation} \sigma_{{}_{\mathit{KK}}}(pp
\rightarrow  g^{\star} g^{\star} \rightarrow \, 4 \,
\mathrm{jets}) = \int \frac{d\Omega_5}{4\pi} \int
\frac{d\Omega_7}{4\pi} \sigma_{{}_{\mathit{KK}}}(pp \rightarrow
g^{\star}g^{\star}) \, .
\end{equation}

\begin{figure}
\setlength{\abovecaptionskip}{0pt}
\centering{\includegraphics[bb=144 202 468 676]{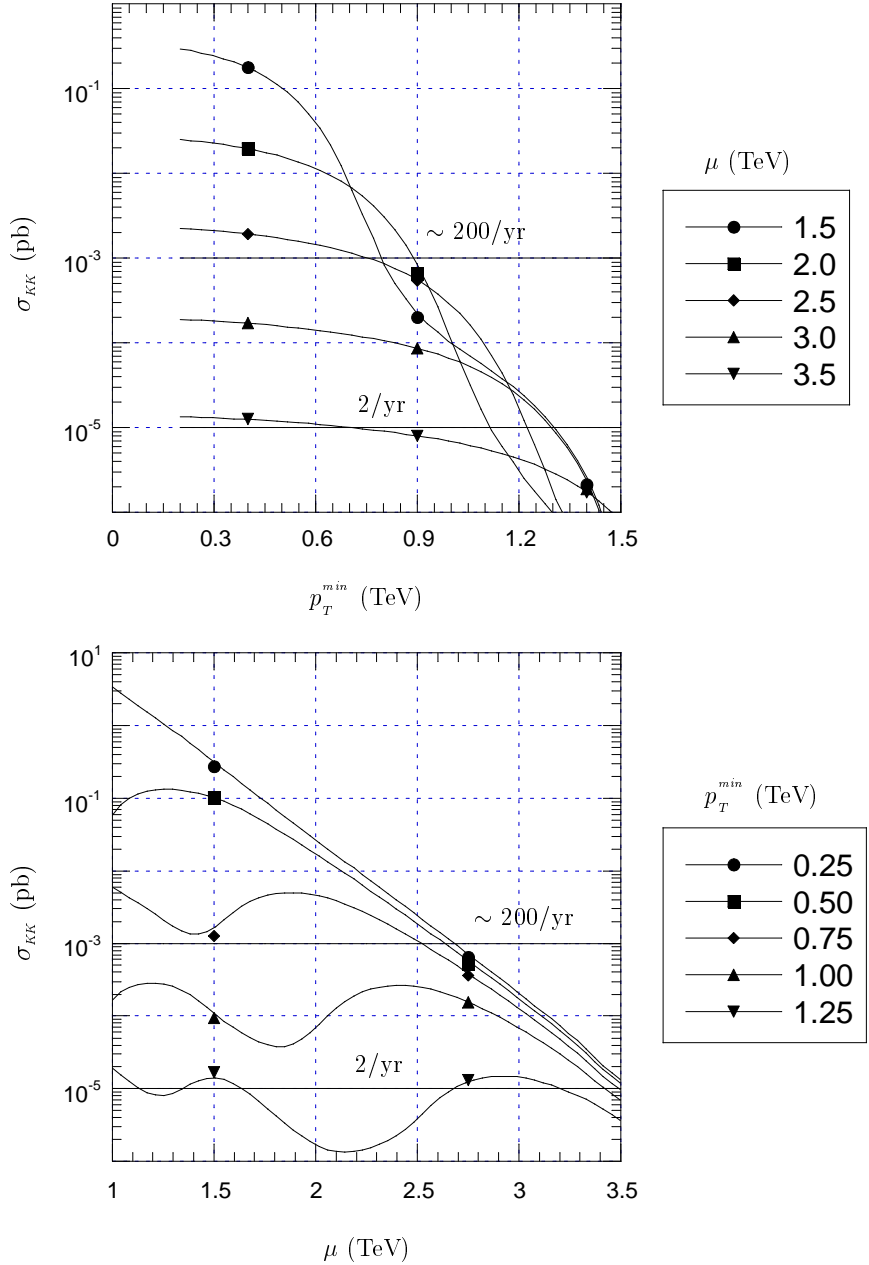}}
\vspace{10pt} \caption{The contributions of the double-on shell
production of $g^{\star}$'s to the four-jet cross section at the
LHC, $\sigma_{{}_{\mathit{KK}}}  =  \sigma -
\sigma_{{}_{\mathit{SM}}}$, are illustrated as a function of the
minimum transverse momentum $p_{{}_T}^{{}_{\mathit{min}}}$ for
fixed values of the compactification scale $\mu$ (top) and as a
function of $\mu$ for fixed $p_{{}_T}^{{}_{\mathit{min}}}$
(bottom).}\label{fig:gs4m} \setlength{\abovecaptionskip}{0pt}
\end{figure}

\noindent We employ the same cuts utilized for the single
$g^{\star}$ case. Illustrated in Fig.~\ref{fig:gs4m}
are the four-jet KK cross sections for $1.0$ TeV $\le \mu \le 3.5$
TeV , and $p_{{}_T}^{{}^{\mathit{min}}}  \leq  1.5$ TeV.  High
$p_{{}_T}$ cuts have a similar effect to that described for single
$g^{\star}$ production. The KK cross section is considerably
smaller for double $g^{\star}$ production as compared to the
single $g^{\star}$ case, which itself is much smaller than the
dijet case:  For double $g^{\star}$ production, the KK cross
section is too small to expect more than a couple of events per
year for a compactification scale in excess of $3.5$ TeV,
regardless of the SM four-jet background. The subprocess with
initial quarks is about a factor of $6$ larger than the
contribution from initial gluons, which can be explained by the
fact that it is partially magnified by the factors of $\sqrt{2}$
in the $q$-$\bar{q}$-$g^{\star}$ vertices.  Also, the production
of two $g^{\star}$'s with different modes is negligible compared
to the case when they have identical modes because there can not
be a gluon propagator in the s-channel in the former case.

\vspace{0.5cm}

\noindent
{\bf 6.  Conclusions}

\vspace{0.2cm}

\noindent In this work, we have investigated the phenomenology of
a class of string-inspired models in which the SM gauge bosons can
propagate into one TeV-scale extra dimension.  Specifically, we
calculate the effects that the KK excitations of the gluons have
on multijet final states at very high energy hadronic colliders
such as the LHC or upgraded Tevatron Run 2.

At the LHC ($\sqrt{s} = 14$ TeV), we found a large enhancement,
relative to the SM, of the dijet cross sections at high
$p_{{}_T}$, while at the upgraded Tevatron we found an effect that
is considerably smaller.  The effect is observable at the LHC for
a compactification scale $\mu \lesssim 7$ TeV, for a wide range of
very high $p_{{}_T}$.  For example, with a minimum $p_{{}_T}$ for
each of the jets of $2$ TeV, the dijet cross section is about
three times larger than that of the SM for $\mu = 5$ TeV.  Thus,
the measurements of the dijet cross sections at the LHC will
either discover the indirect effects of the KK modes of the gluons
or set a bound on $\mu$ of about $7$ TeV, which is significantly
higher than the current bound of about $2$ TeV. The effect is much
less discernible at the upgraded Tevatron, and will not be
observed for $\mu \gtrsim 2$ TeV.  For three jets in the final
state, in which two of the jets are the decay products of an
on-shell $g^{\star}$, at high $p_{{}_T}$ at the LHC, the KK
enhancement over the SM cross sections is much smaller than for
the dijet case. For example, with a minimum $p_{{}_T}$ of each of
the jets of $1.5$ TeV, the cross section is enhanced only by about
$100\%$ for $\mu = 3$ TeV. Although the dijet effect is much
greater, three-jet final state measurements can offer additional
confirming information if a large effect is seen in dijet final
state measurements.  For four jets in the final state from double
on-shell $g^{\star}$ production, again the cross sections are
rather small unless $\mu \lesssim 2.5$ TeV.

In the case of single or double on-shell $g^{\star}$ production
leading to three or four jets, respectively, in the final state,
the on-shell $g^{\star}$'s subsequently decay primarily (the
exceptions involve loop corrections) to quark and anti-quark
pairs.  These quark and anti-quark decay products will have very
high $p_{{}_T}$ because the mass of the $g^{\star}$ is quite high
(some multiple of the compactification scale, which is at least a
TeV).  If the invariant mass of the parent particle can be
reconstructed using the measured high $p_{{}_T}$ of the jets, then
that will be the clear signal of the first KK excitation of the
gluons.  In the three-jet case, such reconstruction must be done
for each pair-wise configuration.  Thus, for three jets in the
final state, although the total cross section is not much larger
than the SM background, such an invariant mass peak could
potentially stand well above the SM background.

Now, we discuss some of the uncertainties in our calculations and
results.  Firstly, in the parton distribution function
$f_{a/{}_A}(x_{{}_A},Q)$ and the strong coupling $\alpha_{{}_S}
(Q)$, our results are somewhat sensitive to the choice of the
scale $Q$.  We chose $Q = p_{{}_T}$ for the SM background as well
as for the KK contribution to the dijet signal, and $Q = m_n$
(\textit{i.e.}, the mass of the $g^{\star}$) for single and double
$g^{\star}$ production. We varied $Q$ from $p_{{}_T}/2$ to $2
p_{{}_T}$ for two or three jets in the final state for the SM
background, and found an enhancement of about $40\%$ for
$p_{{}_T}/2$ and a reduction of about $30\%$ for $2 p_{{}_T}$
compared to $Q = p_{{}_T}$.  Thus, if the KK effect does not
exceed the SM background significantly, it may be difficult to
discern in light of the uncertainty arising from the choice of
$Q$. However, for two jets only, we employ the same value for $Q$
in the KK and SM cases, such that this uncertainty has less
relative effect on the ratio $R$.  Therefore, $R$ can be somewhat
smaller for two jets than three or more jets and still provide
indirect evidence of KK excitations of the gluons.  Secondly, in
the calculations of three- and four-jet cross sections, we have
only considered the production of $g^{\star}$'s on-shell and their
subsequent decays. We have not included those diagrams involving
virtual $g^{\star}$'s.  Such virtual $g^{\star}$ contributions
will naturally be small because they are higher order in the
strong coupling constant $\alpha_{{}_S} (Q)$. However, there are
many virtual $g^{\star}$ diagrams (especially for four-jet
diagrams) which may lead to a sizeable total contribution.
Inclusion of these virtual $g^{\star}$ diagrams would enhance our
three- and four-jet signals, thereby producing a somewhat greater
effect. Finally, we have evaluated the running of the strong
coupling constant $\alpha_{{}_S} (Q)$ with the usual logarithmic
behavior of the SM. This is fine for $Q \le \mu$, but when $Q >
\mu$, the decrease is a power law behavior, in which case
$\alpha_{{}_S} (Q)$ would be somewhat smaller. However, since in
most of our calculations, the scale $Q$ (which is equal to
$p_{{}_T}$ in the dijet case and $m_n$ otherwise) is less than
$\mu$ or does not exceed $\mu$ by much, the net effect would be
only a relatively small reduction of our calculated cross sections
(in our scenario with only one extra dimension).

Finally, we address the issue of how to distinguish the signal due
to KK excitations from other new physics that might produce a
similar collider signal.  For example, the colorons \cite{coloron}
in the top color model produce effects similar to those of the KK
excitations of the gluons. The eight colorons are like eight heavy
gluons with the same mass, whereas, in the KK case, there is an
infinite tower of increasing masses, $m_n = n \mu$ ($n =
1,2,\ldots$).  One important distinguishing feature between the
two cases is the difference in the details of the decay modes of
the colorons and the KK excitations of the gluons.  While the
branching ratios of the KK $g^{\star}$'s to the various quark
flavors are identical, the branching ratios of the coloron to
various flavors of quarks ($q_i \bar{q}_i$, $i \in
\{u$,$d$,$c$,$s$,$t$,$b\}$) depends on the mixing angle between
the two SU(3)'s, SU(3)$_I$ and SU(3)$_{II}$.  In the limit of zero
mixing angle, the colorons couple only to $t \bar{t}$ and $b
\bar{b}$.  Thus, while the KK $g^{\star}$'s decay equally to
various quark flavors, the coloron decay is flavor-dependent.  In
the small mixing case, the dominant decays will be to $t \bar{t}$
and $b \bar{b}$.  For the $t \bar{t}$ decay, the $p_T$ of the jets
coming from the subsequent decay of the top quark will be reduced.
Thus, the dijet signal at very high $p_{{}_T}$ would be much
stronger in the KK case than in the coloron case.

\vspace{0.5cm}



\noindent We are grateful to K.S. Babu, J.D. Lykken, and J.D.
Wells for useful discussions.  The work of DD was supported in
part by the U.S. Department of Energy Grant Number
DE-FG03-93ER40757; the work of CM and SN by DE-FG03-98ER41076.

\vspace{0.5cm}

\noindent
{\bf Appendix
\appendix}

\vspace{0.2cm}

\noindent The generalization of the $4$D SM Lagrangian density to
the $5$D Lagrangian density leads to $5$D gluon field strength
tensors $F_{\mathit{MN}}^a  =  \partial_M A_N^a - \partial_N A_M^a
- g_{{}_5} f^{\mathit{abc}} A_{\mathit{M}}^b A_{\mathit{N}}^c$
described by

\pagebreak[2] \vspace{-8pt} \begin{eqnarray} \label{eq:L5}
\mathcal{L}_5  =  & \!\!\! - & \!\!\!\!\! \frac{1}{4} F_{\mathit{MN}}^a F^{\mathit{MNa}}  +  i \bar{q} \gamma^{\mu} D_{\mu} q \delta (y) \nonumber \\
= & \!\!\! - & \!\!\!\!\! \frac{1}{4}\mbox{\raisebox{-.6ex}{\huge $($}}F_{\mu \nu}^a F^{\mu \nu a}  +  2 F_{\mu 4}^a F^{\mu 4 a}\mbox{\raisebox{-.6ex}{\huge $)$}} \nonumber \\
& \!\!\! + & \!\!\! i \bar{q} \gamma^{\mu} D_{\mu} q \delta (y) \,
,
\end{eqnarray}

\noindent where $g_{{}_5}$ is the $5$D strong coupling, $A_M^a$ is
the $5$D gluon field, $a$,$b$,$c$ are the usual gluon color
indices, $D_{\mu}$ is the usual $4$D covariant derivative,
$\mu$,$\nu$ are the usual $4$D space-time indices, $M$,$N  \in
\{0,1,...,4\}$ are $5$D space-time indices, and $\delta(y)$
represents that the SM fermions are localized in the D$_3$ brane
with $y  =  0$. The terms representing the kinetic energy and
interactions between the $g$ and $g^{\star}$ fields arise from the
contraction of the $F_{\mu \nu}^a$'s:

\vspace{-8pt} \begin{eqnarray} F_{\mu \nu}^a F^{\mu \nu a} = &
\!\!\! & \!\!\!\!\!\!\!\!\!\!
\partial_{\mu} A_{\nu}^a
\partial^{\mu} A^{\nu a} \,
- \, \partial_{\nu} A_{\mu}^a \partial^{\mu} A^{\nu a} \nonumber \\
& \!\!\! - & \!\!\! \partial_{\mu} A_{\nu}^a \partial^{\nu} A^{\mu a}  +  \partial_{\nu} A_{\mu}^a \partial^{\nu} A^{\mu a} \nonumber \\
& \!\!\! - & \!\!\! 2 g_{{}_5} f^{\mathit{abc}} A_{\mu}^b A_{\nu}^c \mbox{\raisebox{-.6ex}{\huge $($}}\partial^{\mu} A^{\nu a}  -  \partial^{\nu} A^{\mu a}\mbox{\raisebox{-.6ex}{\huge $)$}} \nonumber \\
& \!\!\! - & \!\!\! g_{{}_5}^2 f^{\mathit{abc}} f^{\mathit{ade}}
A_{\mu}^b A_{\nu}^c A^{\mu d} A^{\nu e} \, .
\end{eqnarray}

\noindent Similarly, the mass terms for the $g_n^{\star}$'s stem
from the contraction of the $F_{\mu 4}^a$'s:

\vspace{-8pt} \begin{equation} F_{\mu 4}^a F^{\mu 4 a} =
\partial_4 A_{\mu}^a
\partial^4 A^{\mu a} \, ,
\end{equation}

\noindent where the gauge choice $A_{4}^a  =  0$ has been imposed.
The remaining interaction of the $g^{\star}$'s involves the quark
fields and is governed by the term in Eq.~(\ref{eq:L5}) involving
the covariant derivative. We consider compactification on a $S^1 /
Z_2$ orbifold and make the identification $y  \rightarrow -y$ such
that $A_{\mu}^a (x,-y) = A_{\mu}^a (x,y)$. The fields $A_{\mu}^a
(x,y)$ can then be Fourier expanded in terms of the compactified
dimension $y  =  r \phi$ as

\vspace{-3pt} \begin{equation} A_{\mu}^a (x,y)  = \frac{1}{\sqrt{
\pi r}}\mbox{\raisebox{-.6ex}{\huge $[$}}A_{\mu 0}^a (x) +
\sum_{n=1}^{\infty}A_{\mu n}^a (x)
\cos(n\phi)\mbox{\raisebox{-.6ex}{\huge $]$}} \, ,
\end{equation}

\noindent
where the normalization of $A_0^a (x)$ for the gluon field is one-half that of $A_n^a (x)$
for the KK excitations.

Integration over the compactified dimension $y$ then gives the effective $4$D
theory.  The terms from the integration of
$-\frac{1}{4}F_{\mu \nu}^a F^{\mu \nu a}$ over $y$ that are
quadratic in the fields $A_{\mu}^a (x,y)$ give rise to kinetic energy terms in the effective
$4$D Lagrangian density of the form

\vspace{-8pt} \begin{eqnarray}
-\frac{1}{4}\int_{\mbox{\raisebox{-1.3ex}{\scriptsize{$\!\!\!\!
0$}}}}^{\mbox{\raisebox{.9ex}{\scriptsize{$\!\!\!\! \pi r$}}}}
\partial_{\mu} A_{\nu}^a (x,y)
\partial^{\mu} A^{\nu a} (x,y) dy = &\!\!\! -& \!\!\!\!\!\frac{1}{4}
\mbox{\raisebox{-.6ex}{\huge $[$}}\partial_{\mu} A_{\nu 0}^a (x)
\partial^{\mu} A_0^{\nu a} (x) \nonumber \\
&\!\!\! + &\!\!\! \frac{1}{2} \sum_{n=1}^{\infty} \partial_{\mu}
A_{\nu n}^a (x) \partial^{\mu} A_n^{\nu a}
(x)\mbox{\raisebox{-.6ex}{\huge $]$}} \, .
\end{eqnarray}

\noindent
It is then necessary to rescale the fields as

\vspace{-3pt} \begin{equation} A_{\mu 0}^a (x)  \rightarrow  A{}^{
' }{}_{\mu 0}^a (x) , A_{\mu n}^a (x)  \rightarrow  A{}^{ '
}{}_{\mu n}^a (x) \equiv \frac{A_{\mu n}^a (x)}{\sqrt{2}}
\end{equation}

\noindent in order to canonically normalize the kinetic energy
terms. Therefore, the mass and interaction terms must be expressed
in terms of the rescaled fields, $A{}^{ ' }{}_{\mu 0}^a (x)$ and
$A{}^{ ' }{}_{\mu n}^a (x)$.  The masses of the KK excitations of
the gluons arise from the integration of $F_{\mu 4}^a F^{\mu 4 a}$
over $y$:

\vspace{-3pt} \begin{equation}
-\frac{1}{4}\int_{\mbox{\raisebox{-1.3ex}{\scriptsize{$\!\!\!\!
0$}}}}^{\mbox{\raisebox{.9ex}{\scriptsize{$\!\!\!\! \pi r$}}}}
\partial_{4} A_{\mu}^a (x,y)
\partial^{4} A^{\mu a} (x,y) dy  =
-\frac{1}{2}\frac{n^2}{r^2}\sum_{n=1}^{\infty}A{}^{ ' }{}_{\mu
n}^a (x) A{}^{ ' }{}_n^{\mu a} (x) \, .
\end{equation}

\noindent The mass of the $g_n^{\star}$ is then identified as $m_n
=  n \mu$, where $\mu$ is the compactification scale ($\mu  =
1/r$).

The Feynman rules for vertices involving $g^{\star}$'s follow from
the interaction terms.  The interactions of the $g^{\star}$'s with
the quark fields originate from the term in the $5$D Lagrangian
density involving the covariant derivative.  The delta function,
which constrains the quark fields to the wall, takes care of the
integration.  Thus, the $q$-$\bar{q}$-$g^{\star}$ vertex receives
a factor of $\sqrt{2}$, compared to the SM $q$-$\bar{q}$-$g$
vertex, from the rescaling of the $A_{\mu n}^a$ field:

\vspace{-8pt} \begin{equation}
-i\Lambda_{q\mbox{-}\bar{q}\mbox{-}g^{\star}}  =  - i \sqrt{2}
\Lambda_{q\mbox{-}\bar{q}\mbox{-}g} \, ,
\end{equation}

\noindent where the $4$D strong coupling constant $g $ is related
to $g_{{}_5}$ by $g  \equiv  g_{{}_5} / \sqrt{\pi r}$.
Interactions between $g$'s and $g^{\star}$'s are somewhat more
involved. The cubic interaction terms in the effective $4$D
Lagrangian density are

\vspace{-8pt} \begin{eqnarray} - i
\frac{1}{2}&\!\!\!\!\!\!\!g_{{}_5} &\!\!\!\!\!\! f^{\mathit{abc}}
\int_{\mbox{\raisebox{-1.3ex}{\scriptsize{$\!\!\!\!
0$}}}}^{\mbox{\raisebox{.9ex}{\scriptsize{$\!\!\!\! \pi r$}}}} A_{\mu}^b (x,y) A_{\nu}^c (x,y) \mbox{\raisebox{-.6ex}{\huge $[$}}\partial^{\mu}A^{\nu a} (x,y)  -  \partial^{\nu} A^{\mu a} (x,y)\mbox{\raisebox{-.6ex}{\huge $]$}} d y \nonumber \\
= & \!\!\! -&\!\!\!\!\!\frac{1}{2} g f^{\mathit{abc}} \mbox{\raisebox{-.6ex}{\huge $\{$}}A{}^{ ' }{}_{\mu 0}^b (x) A{}^{ ' }{}_{\nu 0}^c (x) \mbox{\raisebox{-.6ex}{\huge $[$}}\partial^{\mu}A{}^{ ' }{}_0^{\nu a} (x)  -  \partial^{\nu} A{}^{ ' }{}_0^{\mu a} (x)\mbox{\raisebox{-.6ex}{\huge $]$}} \nonumber \\
& \!\!\! + & \!\!\! 3 A{}^{ ' }{}_{\mu 0}^b (x) \sum_{n=1}^{\infty} A{}^{ ' }{}_{\nu n}^c (x) \mbox{\raisebox{-.6ex}{\huge $[$}}\partial^{\mu}A{}^{ ' }{}_n^{\nu a} (x)  -  \partial^{\nu} A{}^{ ' }{}_n^{\mu a} (x)\mbox{\raisebox{-.6ex}{\huge $]$}} \nonumber \\
& \!\!\! + & \!\!\! \frac{1}{\sqrt{2}} \sum_{n,m,\ell =1}^{\infty}
A{}^{ ' }{}_{\mu n}^b (x) A{}^{ ' }{}_{\nu m}^c
\mbox{\raisebox{-.6ex}{\huge $[$}}
\partial^{\mu}A{}^{ ' }{}_{\ell}^{\nu a} (x)  -  \partial^{\nu}
A{}^{ ' }{}_{\ell}^{\mu a} (x)\mbox{\raisebox{-.6ex}{\huge $]$}}
\delta_{\ell, \pm m \pm n} \mbox{\raisebox{-.6ex}{\huge $\}$}} \,
,
\end{eqnarray}

\noindent where we introduce the following notation:  The
Kronecker $\delta$ with $\pm$'s represents the summation over all
of the Kronecker $\delta$'s that can be constructed by permuting
the $+$ and $-$ signs (\textit{e.g.}, $\delta_{\ell, \pm m \pm n}
= \delta_{\ell, m+n}  +  \delta_{\ell, m-n}  + \delta_{\ell, n-m}
 +  \delta_{\ell, -m-n}$). These cubic interaction terms lead
to the following Feynman rules for triple vertices involving $g$'s
and $g^{\star}$'s:

\pagebreak[2] \vspace{-8pt} \begin{eqnarray}
-i\Lambda_{g\mbox{-}g_n^{\star}\mbox{-}g_n^{\star}} &\!\!\! = &\!\!\! - i \Lambda_{g\mbox{-}g\mbox{-}g} \nonumber \\
-i\Lambda_{g_n^{\star}\mbox{-}g_n^{\star}\mbox{-}g_{2n}^{\star}} &\!\!\! = &\!\!\! - i \frac{1}{2}\Lambda_{g\mbox{-}g\mbox{-}g} \\
-i\Lambda_{g_n^{\star}\mbox{-}g_m^{\star}\mbox{-}g_{\mid m \pm n
\mid}^{\star}} &\!\!\! = &\!\!\! - i
\frac{1}{2}\Lambda_{g\mbox{-}g\mbox{-}g} \, , \nonumber
\end{eqnarray}

\noindent
for $n \neq m$.  Similarly, the quartic interaction terms in the effective $4$D Lagrangian density are

\vspace{-8pt} \begin{eqnarray} - \frac{1}{4}&\!\!\!\!\!\!\!
g_{{}_5}^2 &\!\!\!\!\!\! f^{\mathit{abc}} f^{\mathit{ade}}
\int_{\mbox{\raisebox{-1.3ex}{\scriptsize{$\!\!\!\!
0$}}}}^{\mbox{\raisebox{.9ex}{\scriptsize{$\!\!\!\! \pi r$}}}} A_{\mu}^b (x,y) A_{\nu}^c (x,y) A^{\mu d} (x,y) A^{\nu e} (x,y) d y \nonumber \\
= &\!\!\! - &\!\!\!\!\! \frac{1}{4} g^2 f^{\mathit{abc}} f^{\mathit{ade}} \mbox{\raisebox{-.6ex}{\huge $[$}}A{}^{ ' }{}_{\mu 0}^b (x) A{}^{ ' }{}_{\nu 0}^c (x) A{}^{ ' }{}_0^{\mu d} (x) A{}^{ ' }{}_0^{\nu e} (x) \nonumber \\
& \!\!\! + & \!\!\! 6 A{}^{ ' }{}_{\mu 0}^b (x) A{}^{ ' }{}_{\nu 0}^c (x) \sum_{n=1}^{\infty} A{}^{ ' }{}_n^{\mu d} (x) A{}^{ ' }{}_n^{\nu e} (x) \nonumber \\
& \!\!\! + & \!\!\! \frac{2}{\sqrt{2}} A{}^{ ' }{}_{\mu 0}^b (x) \sum_{n,m,\ell=1}^{\infty} A{}^{ ' }{}_{\nu n}^c (x) A{}^{ ' }{}_m^{\mu d} (x) A{}^{ ' }{}_{\ell}^{\nu e} (x) \delta_{\ell,\pm m \pm n} \nonumber \\
& \!\!\! + & \!\!\! \frac{1}{2} \sum_{n,m,\ell,k=1}^{\infty} A{}^{
' }{}_{\mu n}^b (x) A{}^{ ' }{}_{\nu m}^c (x) A{}^{ '
}{}_{\ell}^{\mu d} (x) A{}^{ ' }{}_k^{\nu e} (x) \delta_{k,\pm m
\pm n \pm \ell} \mbox{\raisebox{-.6ex}{\huge $]$}} \, .
\end{eqnarray}

\noindent
The Feynman rules for quadruple vertices involving KK excitations are then

\vspace{-8pt} \begin{eqnarray}
-i\Lambda_{g\mbox{-}g\mbox{-}g_n^{\star}\mbox{-}g_n^{\star}} & \!\!\! = & \!\!\! - i \Lambda_{g\mbox{\mbox{-}}g\mbox{-}g\mbox{-}g} \nonumber \\
-i\Lambda_{g\mbox{-}g_n^{\star}\mbox{-}g_n^{\star}\mbox{-}g_{2n}^{\star}} & \!\!\! = & \!\!\! - i \frac{1}{\sqrt{2}} \Lambda_{g\mbox{-}g\mbox{-}g\mbox{-}g} \nonumber \\
-i\Lambda_{g\mbox{-}g_n^{\star}\mbox{-}g_n^{\star}\mbox{-}g_{\mid m \pm n \mid}^{\star}} & \!\!\! = & \!\!\! - i \frac{1}{\sqrt{2}} \Lambda_{g\mbox{-}g\mbox{-}g\mbox{-}g} \nonumber \\
-i\Lambda_{g_n^{\star}\mbox{-}g_n^{\star}\mbox{-}g_n^{\star}\mbox{-}g_n^{\star}} & \!\!\! = & \!\!\! - i \frac{3}{2} \Lambda_{g\mbox{-}g\mbox{-}g\mbox{-}g} \nonumber \\
-i\Lambda_{g_n^{\star}\mbox{-}g_n^{\star}\mbox{-}g_n^{\star}\mbox{-}g_{3n}^{\star}} & \!\!\! = & \!\!\! - i \frac{1}{2} \Lambda_{g\mbox{-}g\mbox{-}g\mbox{-}g} \\
-i\Lambda_{g_n^{\star}\mbox{-}g_n^{\star}\mbox{-}g_m^{\star}\mbox{-}g_m^{\star}} & \!\!\! = & \!\!\! - i \Lambda_{g\mbox{-}g\mbox{-}g\mbox{-}g} \nonumber \\
-i\Lambda_{g_n^{\star}\mbox{-}g_n^{\star}\mbox{-}g_m^{\star}\mbox{-}g_{\mid 2n \pm m \mid}^{\star}} & \!\!\! = & \!\!\! - i \frac{1}{2} \Lambda_{g\mbox{-}g\mbox{-}g\mbox{-}g} \nonumber \\
-i\Lambda_{g_n^{\star}\mbox{-}g_m^{\star}\mbox{-}g_{\ell}^{\star}\mbox{-}g_{\mid
\ell \pm m \pm n \mid}^{\star}} & \!\!\! = & \!\!\! - i
\frac{1}{2} \Lambda_{g\mbox{-}g\mbox{-}g\mbox{-}g} \, , \nonumber
\end{eqnarray}

\noindent
for $n \neq m \neq \ell$.  The relative coupling strengths are summarized
in Fig.~\ref{fig:FDgstar}.
\vspace{0.5cm}

\bibliographystyle{unsrt}

\end{document}